\documentclass[a4paper,11pt]{article}
\usepackage{jheppub}
\usepackage[T1]{fontenc} 
\usepackage{amssymb}
\usepackage{amsmath}
\usepackage{color}
\usepackage{xspace}
\usepackage{caption}
\usepackage{appendix}
\usepackage{float}
\usepackage{hyperref}
\usepackage{diagbox}
\usepackage{soul}
\usepackage{subfig}
\usepackage{slashed}
\usepackage{amsfonts}
\usepackage{multirow}
\usepackage{mathrsfs}
\usepackage{graphicx}
\usepackage{amsmath}
\usepackage{amssymb}
\usepackage{bm}
\usepackage{bbm}
\usepackage{color}
\usepackage{booktabs}

\newcommand{\bqa}{\begin{eqnarray}}
\newcommand{\eqa}{\end{eqnarray}}
\newcommand{\beq}{\begin{equation}}
\newcommand{\eeq}{\end{equation}}
\newcommand{\nn}{\nonumber}

\newcommand{\ep}{\epsilon}

\newcommand{\M}{\ensuremath \text{M}}

\allowdisplaybreaks

\begin{document}

\title{Analytic NNLO QCD corrections to top quark pair production in electron-positron collisions}

\author[b]{Long-Bin Chen,}
\author[c,d]{Jian Wang\footnote{Corresponding author.},}
\author[a]{Yefan Wang\footnote{Corresponding author.}}

\affiliation[a]{Department of Physics and Institute of Theoretical Physics, Nanjing Normal University,
Nanjing, Jiangsu 210023, China}
\affiliation[b]{School of Physics and Materials Science, Guangzhou University, Guangzhou 510006, China}
\affiliation[c]{School of Physics, Shandong University, Jinan, Shandong 250100, China}
\affiliation[d]{Center for High Energy Physics, Peking University, Beijing 100871, China}

\emailAdd{chenlb@gzhu.edu.cn}
\emailAdd{j.wang@sdu.edu.cn}
\emailAdd{wangyefan@nnu.edu.cn}
\vspace{5pt}

\abstract{
We present the analytic total cross section of top quark pair production in electron-positron annihilation at next-to-next-to-leading order (NNLO) in Quantum Chromodynamics (QCD).  By utilizing the optical theorem, the NNLO corrections are related to the imaginary parts of three-loop self-energy Feynman diagrams,
of which the master integrals are calculated with canonical differential equations. The analytic results for the NNLO corrections are expressed in terms of multiple polylogarithms as well as elliptic functions.
We discuss the asymptotic expansions near the threshold and in the high energy limit in detail.
Numerical results are provided for the total cross section of top quark pair production at future lepton colliders.
}

\maketitle

\section{Introduction}

As the heaviest particle in the standard model (SM) of particle physics, the top quark plays a
special role in the precise test of the SM and the exploration of energy frontiers \cite{ParticleDataGroup:2022pth}. 
Since its discovery \cite{CDF:1995wbb,D0:1995jca}, the top quark has attracted continuous research interest.
The top quarks can be copiously produced in pairs via strong interactions at the large hadron collider.
This process provides the dominant background in searching for the signal of new physics, and can also be used to determine the strong coupling and parton distribution functions.
The top quark mass has been measured to be $172.52\pm 0.33$ GeV by utilizing the top quark pair events \cite{ATLAS-CONF-2023-066}. 
The single top quark production is also an important process at the large hadron collider which can be used to extract the CKM matrix element $V_{tb}$ and to set constraints on the anomalous $Wtb$ couplings \cite{Aguilar-Saavedra:2008nuh}.

The future lepton colliders provide more accurate probes to the top quark properties.
The cross section of top quark pair production near the threshold is so sensitive to the top quark mass and width that
 they can be determined with unprecedented precision \cite{Seidel:2013sqa,Horiguchi:2013wra}.
The top quark pair production above the threshold can also be used to measure the top quark mass \cite{Seidel:2013sqa}.
Moreover, the $tt\gamma$ and $ ttZ$ couplings, which get only loose constraints at hadron colliders, can be measured precisely via this process \cite{Janot:2015yza,Amjad:2015mma}, since the top quark pair production is induced by electroweak interaction at lepton colliders.

To match the experimental precision, higher-order QCD and electroweak corrections have to be included. 
The cross section near the top quark pair threshold has been calculated up to QCD next-to-next-to-next-to-leading order (NNNLO) \cite{Czarnecki:1997vz,Beneke:1997jm,Hoang:1998xf,Beneke:1999qg,Beneke:2015kwa}.
The theoretical uncertainty is reduced to $\pm 3\%$.
The cross section above the threshold receives a number of quantum corrections.
The NLO QCD and electroweak corrections were obtained in refs. \cite{Jersak:1981sp,ChokoufeNejad:2016qux} and refs. \cite{Beenakker:1991ca,Fleischer:2003kk}, respectively.
The NNLO QCD corrections were firstly computed by expansion in various limits---the small or heavy quark mass limits and near the threshold---and by Pad\'e  approximants \cite{Chetyrkin:1996cf,Chetyrkin:1997qi,Harlander:1997kw},
and then achieved with full mass dependence by numerical integration methods to properly subtract the infrared divergences \cite{Gao:2014nva,Gao:2014eea,Chen:2016zbz,Bernreuther:2023jgp} or by numerical solution of the differential equations for the three-loop vacuum polarisation diagrams \cite{Maier:2017ypu}.
The flavor non-singlet NNLO contribution was computed analytically in \cite{Dekkers:2014hna} and the flavor non-singlet NNNLO contribution was investigated using the Pad\'e approximation method \cite{Hoang:2008qy,Kiyo:2009gb}.
Recently, the full NNNLO QCD corrections were calculated in the form of piecewise functions which are written as a power series \cite{Chen:2022vzo}.
Discussions on the improvement of renormalization scale dependence using the principle of maximum conformality were carried out in \cite{Yan:2023mjj,Ma:2024izj}.

In this paper, we are going to present the full QCD NNLO analytic result of top quark pair production at a future lepton collider.
This is made possible thanks to the recent improvement in the analytic method of calculating Feynman integrals, especially the proper choice of a canonical basis for the differential equations satisfied by the Feynman integrals \cite{Henn:2013pwa}.
Our analytic result enables an efficient computation of the total cross section of top quark pair production at any collider energy. 
The various limits, such as in the threshold or high energy regions, can be investigated with deep expansion of the full result, which may provide valuable insights into the construction of the corresponding effective theories.
Besides, we provide analytic results for the master integrals with massive propagators, which are useful for understanding the symbol and (pseudo-)pole structure of general loop integrals.

The rest of this paper is organized as follows.
In section \ref{sec:frame}, we set up the relation between the total cross section of $e^-e^+\to t\bar{t}$ and  the forward scattering amplitude of a vector boson, and introduce the coupling structure of top quark pair production
and the workflow of our calculation.
The master integrals needed in this process are computed using the method of differential equations in section \ref{sec:master}.
Then we present analytical results for various contributions to the NNLO QCD corrections in section \ref{sec:nnlo}.
The asymptotic expansions in the threshold and high energy limits are discussed in detail in section \ref{sec:exp},
and the numerical results are exhibited in section \ref{sec:num}.
We conclude in section \ref{sec:conclusion}.

\section{Calculation framework}
\label{sec:frame}

We make use of the optical theorem to calculate QCD higher-order corrections to the total cross section of $e^- e^+\to \gamma^*/Z^*\to t\bar{t}$,
\begin{align}
    \sigma_{e^- e^+\to t\bar{t}} = \frac{1}{4s} {\rm Im} [e^-(p_1) e^+(p_2)\to V^{\mu}(k)\to t\bar{t} \to {V'}^{\nu}(k) \to e^-(p_1) e^+(p_2) ]\,,
\end{align}
where $s=k^2=2p_1 \cdot p_2$ and the factor $1/4$ is due to the polarization average of the initial states, $V$ and $V'$ could be $\gamma$ or $Z$ bosons.
We have taken top quarks as the heavy quarks, but it is ready to extend the results to the other heavy quarks.
The initial- and final-state electron (positron) have the same momenta and spins,
and therefore one can take traces over the lepton fermion chain, obtaining 
\begin{align}
L^{\mu\nu}_{VV'}=c_0^{VV'} (-s g^{\mu\nu}+2p_1^{\mu}p_2^{\nu}+2 p_1^{\nu}p_2^{\mu}),
\end{align}
where
\begin{align}
c_0^{\gamma\gamma} = 8\pi\alpha,\quad
c_0^{ZZ} =  \frac{8\pi\alpha\left(c_{\text{axi}}^2+c_{\text{vec,l}}^2\right)}{\sin^2\theta_W},\quad
c_0^{\gamma Z} =  \frac{8\pi\alpha c_{\text{vec,l}}}{\sin\theta_W}\,
\end{align}
with the couplings between the $Z$ boson and leptons 
\begin{align}
c_{\text{axi}} &= - \frac{1}{4\cos\theta_W},\quad 
c_{\text{vec,l}} = c_{\text{axi}}\left(1-4\sin^2\theta_W\right).
\end{align}
Here $\theta_W$ is the weak mixing angle.
The QCD corrections exist only in the quark sector $V^{\mu}(k)\to t\bar{t} \to {V'}^{\nu}(k)$,
of which the amplitude takes the form
\begin{align}
H_{\mu\nu}^{VV'}=c_1^{VV'} g_{\mu\nu}+c_2^{VV'} k_\mu k_\nu \,.
\end{align}
The second term does not contribute due to the identity $k_{\mu}L^{\mu\nu}_{VV'}=0$.
Then the cross section is given by 
\begin{align}
     \sigma_{e^- e^+\to t\bar{t}} =\sum_{V,V'=\gamma,Z} -\frac{1}{2}c_0^{VV'} {\rm Im}_{t\bar{t}}[ c_1^{VV'}   ]\times \frac{1}{s-m_V^2+im_V\Gamma_V}\frac{1}{s-m_{V'}^2-im_{V'}\Gamma_{V'}}, 
\label{eq:xsection}
\end{align}
where $\Gamma_{V,V'}$ is the decay width of the vector boson and
$ {\rm Im}_{t\bar{t}}[ c_1^{VV'}   ]$  denotes the imaginary part of $c_1^{VV'}$ induced by the cuts on at least a top quark pair.

In the following part, we focus on the calculation of $c_1^{VV'}$,
which can be obtained from $H_{\mu\nu}^{VV'}$ by
\begin{align}
    c_1^{VV'} = \frac{1}{d-1} \left( g^{\mu\nu} H_{\mu\nu}^{VV'} - \frac{1}{s}k^{\mu}k^{\nu} H_{\mu\nu}^{VV'} \right)\,.
    \label{eq:projc1}
\end{align}
Because the intermediate vector bosons $V,V'$ can be either a photon or a $Z$ boson, the cross section is decomposed to three parts,
\begin{align}
\sigma_{e^- e^+\to t\bar{t}} &= \sigma_{e^- e^+\to t\bar{t}}^{\gamma\gamma} +  \sigma_{e^- e^+\to t\bar{t}}^{ZZ}  +  \sigma_{e^- e^+\to t\bar{t}}^{\gamma Z}\,, 
\end{align}
where the superscripts denote the intermediate vector bosons.
We consider QCD higher-order corrections, and expand the cross section in a series of the strong coupling $\alpha_s$,
\begin{align}
\sigma^{V V'}_{e^- e^+\to t\bar{t}} &= \sigma^{V V'}_{\text{LO}}\left(1+\frac{\alpha_s}{\pi}X_1^{V V'}+\left(\frac{\alpha_s}{\pi}\right)^2X_2^{V V'}+\mathcal{O}(\alpha_s^3)\right).
\end{align}
The leading order (LO) results read 
\begin{align}
\sigma^{\gamma\gamma}_{\text{LO}} &= \frac{4\pi \alpha^2C_Ae_u^2\sqrt{s \left(s-4m_t^2\right)} \left(s+2m_t^2\right)}{3s^3},\\
\sigma^{ZZ}_{\text{LO}} &= \frac{4 \pi  \alpha ^2 C_A \left(c_{\text{axi}}^2+c_{\text{vec,l}}^2\right) \sqrt{s \left(s-4 m_t^2\right)} \left(c_{\text{axi}}^2 \left(s-4 m_t^2\right)+c_{\text{vec,u}}^2 \left(s+2 m_t^2\right)\right)}{3\sin^4\theta_W s\left(\left(s-m_Z^2\right)^2+m_Z^2\Gamma_Z^2\right)}
,\label{eq:sigmaZZ}\\
\sigma^{\gamma Z}_{\text{LO}} &= -\frac{8 \pi  \alpha ^2 C_A \,c_{\text{vec,u}} \,c_{\text{vec,l}} \,e_u \sqrt{s \left(s-4 m_t^2\right)} \left(s+2 m_t^2\right)(s-m_Z^2)}{3\sin^2\theta_W s^2 \left(\left(s-m_Z^2\right)^2+m_Z^2\Gamma_Z^2\right)},
\end{align}
where $e_u$ and $m_t$ are the charge and mass of the top-quark, 
$m_Z$  and $\Gamma_Z$ represent the mass and width of the $Z$ boson,
and the couplings between the $Z$ boson and quarks are given by
\begin{align}
c_{\text{vec,u}} & = -c_{\text{axi}}\left(1-\frac{8}{3}\sin^2\theta_W\right), \quad
c_{\text{vec,d}} = c_{\text{axi}}\left(1-\frac{4}{3}\sin^2\theta_W\right).
\end{align}
The NLO QCD corrections were computed in refs. \cite{Jersak:1981sp,ChokoufeNejad:2016qux}.
We have calculated them independently and obtained 
\begin{align}
&X_1^{ZZ} = 
\frac{c_{\text{vec,u}}^2C_{F}}{4 (w-1)^3 (w+1) \left(c_{\text{axi}}^2 (w+1)^2+c_{\text{vec,u}}^2 \left(w^2-4 w+1\right)\right)}\times
\nonumber\\&\quad
\Big[8 (w-1)^2 \left(w^2+1\right) \left(w^2-4 w+1\right) 
\nonumber\\&\quad
\times(-2 G(-1,0,w)+2 G(0,-1,w)+G(0,1,w)-G(1,0,w))
\nonumber\\&\quad
+8 (w-1)^2 \left(w^2-4 w+1\right) \left(i \pi  w^2-w^2+i \pi +1\right) (2 G(-1,w)+G(1,w))
\nonumber\\&\quad
+4 (w-1)^2 w \left(6 w^3-22 w^2+7 w+2\right) G(0,w)
\nonumber\\&\quad
+(w-1)^2 \left(3 (1-8 i \pi ) w^4+(-24+88 i \pi ) w^3-28 i \pi  w^2+8 (3-i \pi ) w-3\right)\Big]
\nonumber\\&\quad
+\frac{c_{\text{axi}}^2C_{F}}{4 (w-1)^3 (w+1) \left(c_{\text{axi}}^2 (w+1)^2+c_{\text{vec,u}}^2 \left(w^2-4 w+1\right)\right)}\times
\nonumber\\
&\quad\Big[8\left(-w^6+w^4+w^2-1\right) G(1,0,w)-8 (w-1)^2 (w+1)^2 \left(w^2+1\right) 
\nonumber\\
&\quad\times(2 G(-1,0,w)-2 G(0,-1,w)-G(0,1,w))
\nonumber\\&\quad 
+8 (w-1)^2 (w+1)^2 \left(i \pi  w^2-w^2+i \pi +1\right) (2 G(-1,w)+G(1,w))
\nonumber\\&\quad
+4 w \left(6 w^5-7 w^4-3 w^3-10 w^2+15 w-7\right) G(0,w)
\nonumber\\&\quad
+3 (1-8 i \pi ) w^6+4 (6+7 i \pi ) w^5+(-45+12 i \pi ) w^4+40 i \pi  w^3
\nonumber\\&\quad
+15 (3-4 i \pi ) w^2+4 i (7 \pi +6 i) w-3\Big],
\end{align}
where the redefinition of the parameter
\begin{align}
	s = -\frac{(w-1)^2}{w}m_t^2
\end{align}
has been used to rationalize the square root $\sqrt{s(s-4m_t^2)}$ with $-1\leqslant w<0$ assuming $s \geqslant 4m_t^2$. The multiple polylogarithms (MPLs) \cite{Goncharov:1998kja} are defined by $G(x)\equiv 1$ and
\bqa
	G(l_1,l_2,\ldots,l_n,x) &\equiv & \int_0^x \frac{\text{d} t}{t - l_1} G(l_2,\ldots,l_n,t)\, ,\\
	G(\overrightarrow{0}_n , x) & \equiv & \frac{1}{n!}\ln^n x\, .
\eqa
The number of elements in the set $\{l_1,l_2,\ldots,l_n\}$ is referred to as the transcendental $weight$ of the MPLs.

One can see that there are both vector and axial-vector  contributions in $X_1^{ZZ}$. 
But $X_1^{\gamma\gamma}$ and $X_1^{\gamma Z}$ only have vector contributions,
\begin{align}
X_1^{\gamma\gamma} = X_1^{\gamma Z} = X_1^{Z Z}\bigg|_{c_{\text{axi}}=0}\,.
\end{align}

\begin{figure}[ht]
	\centering
\begin{minipage}{0.3\linewidth}
	\centering
	\includegraphics[width=0.8\linewidth]{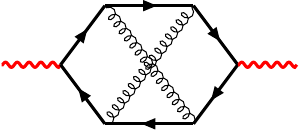}
 \caption*{(a)}
\end{minipage}
 \begin{minipage}{0.3\linewidth}
		\centering
		\includegraphics[width=0.8\linewidth]{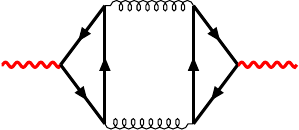}
   \caption*{(b)}
\end{minipage}
\begin{minipage}{0.3\linewidth}
		\centering
		\includegraphics[width=0.8\linewidth]{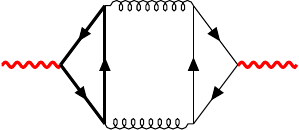}
  \caption*{(c)} 
\end{minipage}
\caption{Typical three-loop Feynman diagrams of $ V \rightarrow t \bar{t} \rightarrow V' $ contributing to $X_{2,Vt\Bar{t}}$. The thick black and red wavy lines stand for the massive top quark and the vector boson, respectively. The thin lines
represent the massless fermions. The curly lines denote gluons.  }
\label{ThreeLoop_Vtt}
\end{figure}

\begin{figure}[ht]
	\centering
	\begin{minipage}{0.4\linewidth}
		\centering
		\includegraphics[width=0.9\linewidth]{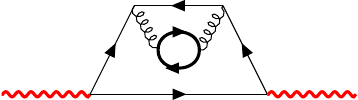}
	\end{minipage}
	\begin{minipage}{0.4\linewidth}
		\centering
		\includegraphics[width=0.8\linewidth]{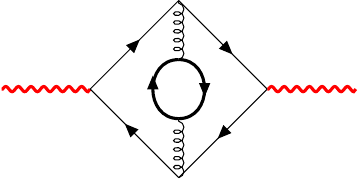}
	\end{minipage}
\caption{Typical three-loop Feynman diagrams of $V \rightarrow t \bar{t} \rightarrow V'$ contributing to $X_{2,gt\Bar{t}}$. The thick black and red wavy lines stand for the massive top quark and the vector boson, respectively. The thin lines
represent the massless fermions. The curly lines denote gluons.  }
\label{ThreeLoop_gtt}
\end{figure}

To calculate the NNLO QCD corrections, we need to consider the three-loop Feynman diagrams of $V\to t\bar{t} \to V'$.
Some typical diagrams are shown in figure \ref{ThreeLoop_Vtt} and figure \ref{ThreeLoop_gtt}.
Figure \ref{ThreeLoop_Vtt}(a) shows the flavor non-singlet contribution, which is induced either by a vector or axial-vector current.
Figure \ref{ThreeLoop_Vtt}(b) and (c) represent the flavor singlet contribution, in which only axial-vector currents give non-vanishing results.
In the diagrams in figure \ref{ThreeLoop_Vtt}, the top quark couples to the electroweak boson, and thus these contributions are denoted by $X_{2,Vt\bar{t}}$.
In the diagrams in figure \ref{ThreeLoop_gtt}, the top quark does not enter the electroweak current directly but arises from a gluon splitting, and the corresponding contributions are labelled by $X_{2,gt\bar{t}}$.
Notice that diagrams in figure \ref{ThreeLoop_Vtt}(b) and (c) and figure \ref{ThreeLoop_gtt} start from NNLO QCD corrections.

We have generated all the Feynman diagrams and amplitudes of $V^{\mu}\to t\bar{t} \to {V'}^{\nu}$ by FeynArts \cite{Hahn:2000kx},
and simplified the Dirac matrices with FeynCalc \cite{Shtabovenko:2020gxv,Shtabovenko:2023idz}.
We work in $(4-2\ep)$-dimensional space-time to regularize the infrared and ultraviolet divergences.
The treatment of $\gamma_5$ in dimensional regularization is subtle. 
For the flavour non-singlet diagrams, such as the one in figure \ref{ThreeLoop_Vtt}(a), we simply use the naive anticommutative $\gamma_5$ scheme.
The fermion chain with two $\gamma_5$ matrices is easy to deal with since they can be moved together and contracted by $\gamma_5^2=1$.
The fermion chain with a single $\gamma_5$ matrix needs a specific prescription to evaluate the trace generally.
In our case, the trace must be proportional to $\epsilon^{\mu\nu\rho\sigma}\times (d_1k_{\rho}k_{\sigma}+d_2 g_{\rho\sigma})$ after loop momentum integrations and thus vanishes, irrespective of the values of the coefficients $d_1,d_2$.
For the flavour singlet diagrams, such as the one in figure \ref{ThreeLoop_Vtt}(b),
we adopt the Larin scheme to deal with the $\gamma_5$ matrix \cite{Larin:1991tj,Larin:1993tq} and substitute 
\begin{align}
    \gamma^{\mu}\gamma_5 = \frac{i}{3!}\epsilon^{\mu\nu\rho \sigma}\gamma_{\nu}\gamma_{\rho}\gamma_{\sigma}\,
\end{align}
in a fermion chain.
The Dirac matrices on the right-hand side of the above equation should be taken in $(4-2\ep)$ dimensions.
Notice that since the axial-vector current anomalies in figure \ref{ThreeLoop_Vtt}(b) and (c) appear for the first time at $\mathcal{O}(\alpha_s^2)$, one does not need to consider the (either ultraviolet or finite) renormalization constants of the axial-vector current.

\begin{figure}[ht]
	\centering
\begin{minipage}{0.35\linewidth}
	\centering
	\includegraphics[width=0.8\linewidth]{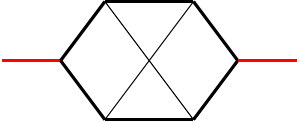}
	\caption*{NP1}
\end{minipage}
\begin{minipage}{0.35\linewidth}
	\centering
	\includegraphics[width=0.8\linewidth]{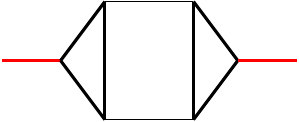}
	\caption*{P1}
\end{minipage}
\begin{minipage}{0.35\linewidth}
	\centering
	\includegraphics[width=0.8\linewidth]{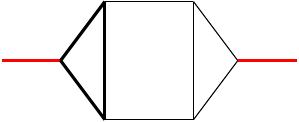}
	\caption*{P2}
\end{minipage}
\begin{minipage}{0.35\linewidth}
	\centering
	\includegraphics[width=0.8\linewidth]{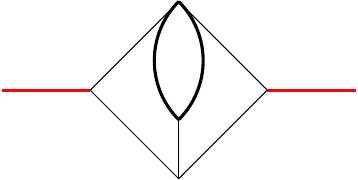}
	\caption*{P3}
\end{minipage}
\caption{The topologies of master integral families. The thick black and red lines stand for the massive top quark and the vector boson, respectively. The other lines represent massless particles.  A cut on at least a top quark pair is assumed.}
\label{MIs_Fam}
\end{figure}

From the amplitudes, we projected out the scalar coefficient $c_1$ making use of eq.~(\ref{eq:projc1}).
Then, all the integrals in $c_1$ were reduced to a set of master integrals using {\tt FIRE} \cite{Smirnov:2019qkx} and {\tt Kira} \cite{Klappert:2020nbg}.
These master integrals belong to four integral families, which are shown in figure \ref{MIs_Fam}.
The analytical results of master integrals in the P1 and NP1 families have been derived in our previous work on $H\rightarrow b\bar{b}$ \cite{Wang:2023xud}.
In the next section, we present the results for the other two integral families.

\section{Calculation of master integrals}
\label{sec:master}

We calculate the master integrals with the method of differential equations \cite{Kotikov:1990kg,Kotikov:1991pm}.
After choosing a proper basis, the differential equation exhibits a canonical form \cite{Henn:2013pwa}, 
i.e., the dimensional parameter $\epsilon$ is factorized out.
Then the solution as a series of $\epsilon$ can be found using iterated integrals.
In some cases, the kinematic dependence in the differential equation can be represented by logarithmic functions, 
and the result is expressed in terms of multiple polylogarithms (MPLs) \cite{Goncharov:1998kja}.
We have obtained the canonical differential equations for the P1 and NP1 families in ref. \cite{Wang:2023xud}. 
Below we show the results for the P2 and P3 families.

\subsection{Canonical basis of the P2 family}
The integrals of the P2  family in figure \ref{MIs_Fam} are defined by 
\begin{align}
	I^{\rm P2}_{n_1,n_2,\ldots,n_{9}}=\textrm{Im}_{t\bar{t}}\int{\mathcal D}^D q_1~{\mathcal D}^D q_2~{\mathcal D}^D q_3\frac{D_9^{-n_9}}{D_1^{n_1}~D_2^{n_2}~D_3^{n_3}~D_4^{n_4}~D_5^{n_5}~D_6^{n_6}~D_7^{n_7}D_8^{n_8}}
\end{align}
with
\begin{align}
	{\mathcal D}^D q_i = \frac{\left(m_t^2 \right)^\epsilon}{i \pi^{D/2}\Gamma(1+\epsilon)}  d^D q_i \ ,\quad D=4-2\epsilon \,,
\end{align}
where all $n_i\geqslant 0,i=1,\cdots, 8$, and $n_9\leqslant 0$. The denominators $D_i$ read
\begin{align}
	D_1 &= q_3^2,&
	D_2 &= (q_3-k)^2,&
	D_3 &= q_1^2-m_t^2,\nonumber\\
	D_4 &= (q_1+k)^2-m_t^2,&
	D_5 &= (q_1-q_2)^2-m_t^2,&
	D_6 &= q_2^2,\nonumber\\
	D_7 &= (q_2+q_3)^2,&
	D_8 &=(q_2+k)^2,&
	D_9 &=(q_1+q_3)^2,
\end{align}
where the Feynman prescription $+i\varepsilon$ has been suppressed. 
The external momentum $k$ satisfies
$k^2 = s$. 
In the above definition of integrals, we have required only the imaginary contribution induced by a cut on at least a top quark pair.

In the P2 integral family, there are sixteen master integrals contributing to $X_{2,Vt\Bar{t}}$. 
To construct the canonical basis, we first select  
\begin{align}
\M^{\text{P2}}_{1}&=\epsilon^3 m_t^2 I^{\text{P2}}_{0,2,2,0,2,0,1,0,0},\quad&
\M^{\text{P2}}_{2}&=\epsilon^3 m_t^2 I^{\text{P2}}_{0,2,2,0,1,0,2,0,0},\quad\nonumber\\
\M^{\text{P2}}_{3}&=(1-2\epsilon)\epsilon m_t^4 I^{\text{P2}}_{0,1,3,0,3,0,1,0,0},\quad&
\M^{\text{P2}}_{4}&=\epsilon^3 m_t^4 I^{\text{P2}}_{2,1,2,1,2,0,0,0,0},\quad\nonumber\\
\M^{\text{P2}}_{5}&=\epsilon^3 m_t^4 I^{\text{P2}}_{2,1,0,2,2,1,0,0,0},\quad&
\M^{\text{P2}}_{6}&=\epsilon^3 m_t^4 I^{\text{P2}}_{2,1,0,2,1,2,0,0,0},\quad\nonumber\\
\M^{\text{P2}}_{7}&=(1-2\epsilon)\epsilon^3 m_t^2 I^{\text{P2}}_{2,0,1,1,2,0,1,0,0},\quad&
\M^{\text{P2}}_{8}&=\epsilon^3 m_t^4 I^{\text{P2}}_{0,2,2,1,0,2,1,0,0},\quad\nonumber\\
\M^{\text{P2}}_{9}&=(1-2\epsilon)\epsilon^3 m_t^2 I^{\text{P2}}_{0,1,0,2,1,2,1,0,0},\quad&
\M^{\text{P2}}_{10}&=(1-2\epsilon)\epsilon^4 m_t^2 I^{\text{P2}}_{1,1,0,2,1,1,1,0,0},\quad\nonumber\\
\M^{\text{P2}}_{11}&=(1-2\epsilon)\epsilon^4 m_t^2 I^{\text{P2}}_{1,1,1,1,2,0,1,0,0},\quad&
\M^{\text{P2}}_{12}&=(1-2\epsilon)\epsilon^3 m_t^4 I^{\text{P2}}_{1,1,2,1,2,0,1,0,0},\quad\nonumber\\
\M^{\text{P2}}_{13}&=(1-2\epsilon)\epsilon^3 m_t^4 I^{\text{P2}}_{1,1,1,1,3,0,1,0,0},\quad&
\M^{\text{P2}}_{14}&=(1-2\epsilon)\epsilon^4 m_t^2 I^{\text{P2}}_{0,2,1,1,1,1,1,0,0},\quad\nonumber\\
\M^{\text{P2}}_{15}&=\epsilon^3 m_t^6 I^{\text{P2}}_{1,2,2,1,0,2,0,1,0},\quad&
\M^{\text{P2}}_{16}&=(1-2\epsilon)\epsilon^4 m_t^4 I^{\text{P2}}_{2,1,1,1,1,1,0,1,0}.
\end{align}
Here all $\M^{\text{P2}}_i$ are dimensionless, and the corresponding topology diagrams are displayed
in figure \ref{P2_Topo} in appendix \ref{appendix:topoMIs}.

Then the canonical basis integrals $F^\text{P2}_{i},~i=1,\dots,16,$ can be constructed as linear combinations of $\M^\text{P2}_{i}$ using a method similar to that in ref. \cite{Argeri:2014qva}. 
Examples can be found in \cite{Chen:2021gjv}.
For simplicity, we define a dimensionless variable $z \equiv s/m_t^2+i0^+$. Then we obtain the following canonical basis of the P2 family: 
\begin{align}
F^{\text{P2}}_1&=\M^{\text{P2}}_1 (-z),\quad F^{\text{P2}}_2=-\left(\M^{\text{P2}}_1+\M^{\text{P2}}_2\right) r_1,\quad 
F^{\text{P2}}_3=-\frac{1}{4} \M^{\text{P2}}_2 (4-z)+\M^{\text{P2}}_3,\quad 
\nonumber\\
F^{\text{P2}}_4&=\M^{\text{P2}}_4 r_1 (-z),\quad 
F^{\text{P2}}_5=\M^{\text{P2}}_5 z^2,\quad 
F^{\text{P2}}_6=\frac{1}{2} \M^{\text{P2}}_5 r_1 z+\M^{\text{P2}}_6 r_1 z,\quad
\nonumber\\
F^{\text{P2}}_7&=-\frac{\M^{\text{P2}}_1 r_1 (7 z-12)}{12 (4-z)}+\frac{\M^{\text{P2}}_2 r_1}{12}-\frac{\M^{\text{P2}}_3 r_1}{3 (4-z)}+\frac{\M^{\text{P2}}_7 r_1}{4-z},\quad 
F^{\text{P2}}_8=\M^{\text{P2}}_8 r_1 (-z),\quad 
\nonumber\\
F^{\text{P2}}_9&=\frac{\M^{\text{P2}}_1 r_1}{3}+\frac{\M^{\text{P2}}_2 r_1 (4-z)}{6 z}-\frac{2 \M^{\text{P2}}_3 r_1}{3 z}-\frac{\M^{\text{P2}}_5 r_1 z}{2}+\M^{\text{P2}}_9 r_1+\M^{\text{P2}}_{10} r_1,\quad 
\nonumber\\
F^{\text{P2}}_{10}&=\M^{\text{P2}}_{10}\,z,\quad
F^{\text{P2}}_{11}=\M^{\text{P2}}_{11} (-z),\quad
\nonumber\\
F^{\text{P2}}_{12}&=\frac{\M^{\text{P2}}_1 r_1}{3}+\frac{\M^{\text{P2}}_2 r_1 (4-z)}{6 z}-\frac{2 \M^{\text{P2}}_3 r_1}{3 z}+\frac{3 \M^{\text{P2}}_{11} r_1}{2}+\M^{\text{P2}}_{12} r_1+\M^{\text{P2}}_{13} r_1,\quad
\nonumber\\
F^{\text{P2}}_{13}&=\M^{\text{P2}}_{13} (-z),\quad
F^{\text{P2}}_{14}=\M^{\text{P2}}_{14} (-z),\quad
F^{\text{P2}}_{15}=\M^{\text{P2}}_{15} r_1 z^2,\quad
F^{\text{P2}}_{16}=\M^{\text{P2}}_{16} z^2,\quad
\end{align}
where
\begin{align}
r_1 = \sqrt{z(z-4)}.
\end{align}
To rationalize the square root $r_1$, we write
\begin{align}
	z = -\frac{(w-1)^2}{w}
\end{align}
with  $-1\leqslant w<0$ assuming $z \geqslant 4$.
This choice of $w$ renders the prescription $w+i0^+$ in analytic continuation. 
Then $r_1$ is rationalized to 
\begin{align}
r_1 = \frac{(w+1) (w-1)}{w}.
\end{align}
The corresponding differential equations for the canonical basis $\boldsymbol{F}^{\text{P2}}$ become 
\begin{align}
\frac{d \boldsymbol{F}^{\text{P2}}}{dw} = \epsilon \left( \sum_{i=1}^3 \frac{\boldsymbol{N_i}^{\text{P2}}}{w-l_i^{\text{P2}}}\right) \boldsymbol{F}^{\text{P2}}
\end{align}
with
\begin{align}
l^{\text{P2}}_1 = 0\,,~ 
l^{\text{P2}}_{2} = 1\,,~
l^{\text{P2}}_{3} = -1\,, 
\end{align}
and $\boldsymbol{N_i}^{\text{P2}}$ being constant matrices.

\subsection{Canonical basis of the P3 family}
The master integrals of the P3 integral family in figure \ref{MIs_Fam} are defined by 
\begin{align}
	I^{\rm P3}_{n_1,n_2,\ldots,n_{9}}=\textrm{Im}_{t\bar{t}}\int{\mathcal D}^D q_1~{\mathcal D}^D q_2~{\mathcal D}^D q_3\frac{D_8^{-n_8}D_9^{-n_9}}{D_1^{n_1}~D_2^{n_2}~D_3^{n_3}~D_4^{n_4}~D_5^{n_5}~D_6^{n_6}~D_7^{n_7}}
\end{align}
with all $n_i\geqslant 0,i=1,\cdots, 8$, and $n_9\leqslant 0$. The denominators $D_i$ read
\begin{align}
	D_1 &= q_1^2,&
	D_2 &= q_2^2,&
	D_3 &= (q_1+q_2)^2,\nonumber\\
	D_4 &= (q_1+q_2-q_3)^2-m_t^2,&
	D_5 &= q_3^2-m_t^2,&
	D_6 &= (q_2-k)^2,\nonumber\\
	D_7 &= (q_1+k)^2,&
	D_8 &=(q_2-q_3)^2,&
	D_9 &=(q_3-k)^2.
\end{align}
As in the case of the P2 family, only the imaginary contributions induced by a cut on at least a top quark pair are required. There are five master integrals in this family. 
To construct the canonical basis, we first select 
\begin{align}
\M^{\text{P3}}_{1}&=\epsilon^3m_t^2I^{\text{P3}}_{0,2,0,2,2,0,1,0,0},\quad&
\M^{\text{P3}}_{2}&=\epsilon^3m_t^2I^{\text{P3}}_{0,2,0,2,1,0,2,0,0},\quad\nonumber\\
\M^{\text{P3}}_{3}&=(1-2\epsilon)\epsilon \,m_t^4I^{\text{P3}}_{0,1,0,3,3,0,1,0,0},\quad&
\M^{\text{P3}}_{4}&=(1-2\epsilon)\epsilon^4 m_t^2 I^{\text{P3}}_{1,1,1,1,1,1,1,0,0},\quad\nonumber\\
\M^{\text{P3}}_{5}&=(1-2\epsilon)\epsilon^3 m_t^4 I^{\text{P3}}_{1,1,1,2,1,1,1,0,0}.\quad&
\end{align}
The corresponding topology diagrams are displayed
in figure \ref{P3_Topo} in appendix \ref{appendix:topoMIs}. 
Then with the help of the  {\tt Libra} package \cite{Lee:2020zfb}, we obtain the following canonical basis of the P3 family: 
\begin{align}
F^{\text{P3}}_1&=\M^{\text{P3}}_1 (-z),\quad F^{\text{P2}}_2=-\left(\M^{\text{P3}}_1+\M^{\text{P3}}_2\right) r_1,\quad 
F^{\text{P3}}_3=-\frac{1}{4} \M^{\text{P2}}_2 (4-z)+\M^{\text{P3}}_3,\quad 
\nonumber\\
F^{\text{P3}}_4&= \M_4^{\text{P3}} z(1-2\epsilon )-4 \M_5^{\text{P3}} z \,\epsilon,\quad
\nonumber\\
F^{\text{P3}}_5&= -\frac{\M_1^{\text{P3}} z \left(6 (z+8) \epsilon ^2+5 z \epsilon +z\right)}{9 r_2 (2 \epsilon +1)}
+\frac{\M_2^{\text{P3}} (z-4) z (6 \epsilon +1)}{18 r_2}-\frac{2 \M_3^{\text{P3}} z \left(12 \epsilon ^2-8 \epsilon -1\right)}{9 r_2 (2 \epsilon +1)}
\nonumber\\&\quad
+\M_4^{\text{P3}} \left(\frac{r_2 z (4 \epsilon -1)}{2 (z+4)}-\frac{z^2 \epsilon }{r_2}\right)
+\frac{\M_5^{\text{P3}} r_2 z (4 \epsilon -1)}{3 (z+4)}
\end{align}
with
\begin{align}
r_2 = \sqrt{z(z+4)}.
\end{align}
Note that the bases $F^{\text{P3}}_1$, $F^{\text{P3}}_2$ and $F^{\text{P3}}_3$ exist also in the P2 family and  that
the bases $F^{\text{P3}}_4$ and $F^{\text{P3}}_5$ are chosen from a reducible sector. 

The analytical results of $F^{\text{P3}}_4$ and $F^{\text{P3}}_5$ depend on two square roots, i.e., $r_1$ and $r_2$. 
Making use of the {\tt RationalizeRoots} package \cite{Besier:2019kco}, we have found a change of the differential variable,
\begin{align}
z = -\frac{\left(y^2+1\right)^2}{(y-1)y (y+1)},  \quad 0<y \leqslant \sqrt{2}-1\,,
\end{align}
so that the above two square roots can be rationalized simultaneously,
\begin{align}
r_1 = \frac{\left(y^2+1\right) \left(y^2+2 y-1\right)}{(y-1) y (y+1)},\quad
r_2 = \frac{\left(y^2+1\right) \left(y^2-2 y-1\right)}{(y-1) y (y+1)}.
\end{align}
The corresponding differential equations for the canonical basis $\boldsymbol{F}^{\text{P3}}$ become 
\begin{align}
\frac{d \boldsymbol{F}^{\text{P3}}}{dy} = \epsilon \left( \sum_{i=1}^9 \frac{\boldsymbol{N_i}^{\text{P3}}}{y-l_i^{\text{P3}}}\right) \boldsymbol{F}^{\text{P3}}
\end{align}
with
\begin{align}
l^{\text{P3}}_1 &= 0\,,~ 
l^{\text{P3}}_{2} = 1\,,~
l^{\text{P3}}_{3} = -1\,,~
l^{\text{P3}}_{4} = i\,,~
l^{\text{P3}}_{5} = -i\,,~\nonumber\\
l^{\text{P3}}_{6} &= \sqrt{2}+1\,,~
l^{\text{P3}}_{7} = -\sqrt{2}+1\,,~
l^{\text{P3}}_{8} = \sqrt{2}-1\,,~
l^{\text{P3}}_{9} = -\sqrt{2}-1\,.~
\end{align}
In the calculations of $X_{2,gt\Bar{t}}$, $F^{\text{P3}}_4$ and $F^{\text{P3}}_5$ are required up to transcendental $weight$ four and three, respectively. Then we find that $l^{\text{P3}}_{i}$, $i = 6,\cdots,9$ do not appear in the expressions of $X_{2,gt\Bar{t}}$.
\subsection{Boundary conditions}

Solving the above differential equations, we get analytical results for the basis integrals in terms of MPLs with undetermined constants.
These constants can be fixed by using the regularity condition of the integrals on pseudo-poles or by
direct analytical calculation at special phase space points.
In our case,  we find that the constants consist of $\zeta(n)$ and $\pi^n$.
On the other hand, high-precision numerical results of the constants can be obtained by comparing the analytical results to the numerical results calculated with the {\tt AMFlow} package \cite{Liu:2017jxz,Liu:2020kpc,Liu:2022chg,Liu:2022mfb}. 
Then the explicit analytical form is reproduced using the PSLQ algorithm \cite{Ferguson:1999aa}.
Specifically, we chose to determine the boundary condition directly in the physical phase space, i.e., $z>4$, and consider the contributions only from cuts on at least a top quark pair.
Therefore, it is unnecessary to consider analytic continuation. 

\begin{figure}[ht]
	\centering
\begin{minipage}{0.35\linewidth}
	\centering
	\includegraphics[width=0.8\linewidth]{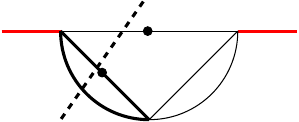}
\end{minipage}
\begin{minipage}{0.35\linewidth}
	\centering
	\includegraphics[width=0.8\linewidth]{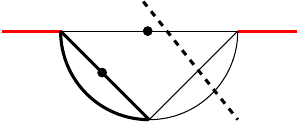}
\end{minipage}
\caption{Cut diagrams of $\M^{\text{P2}}_9$.}
\label{M9_cut}
\end{figure}

For example, the two cut diagrams of $\M^{\text{P2}}_9$ are shown in figure \ref{M9_cut}. The left diagram includes the cut on  $t\bar{t}g$, which is what we need, while the right diagram does not have a cut on the top quark pair. 
We implemented the left cut on the specific propagators in the package {\tt AMFlow} \cite{Liu:2020kpc} to obtain the imaginary part at $z=10$.
Meanwhile, solving the canonical differential equation yields the analytical result. For example, at weight three, the result is 
\begin{align}
&-30 \pi ^2 G(-1,w)+17 \pi ^2 G(0,w)-20 \pi ^2 G(1,w)+36 i \pi  G(-1,-1,w)-30 i \pi  G(-1,0,w)
\nonumber\\
&+24 i \pi  G(-1,1,w)-30 i \pi  G(0,-1,w)+17 i \pi  G(0,0,w)-20 i \pi  G(0,1,w)
+24 i \pi  G(1,-1,w)
\nonumber\\
&-20 i \pi  G(1,0,w)+16 i \pi  G(1,1,w)+C^{\text{P2}}_{9,3}    \,,
\end{align}
where the MPL function can be evaluated using the {\tt PolyLogTools} package \cite{Duhr:2019tlz} and $C^{\text{P2}}_{9,3}$ is an integration constant. 
Comparing the two results,  we obtain
\begin{align}
C^{\text{P2}}_{9,3} =  -356.57218182344793201797762327166i\cdots.
\end{align}
Here the constant at {\it weight} three is imaginary, and it may be a combination of  $i\pi \times \pi^2 $ and $i\pi \times \ln ^2 2 $. A detailed comparison of the digits leads to
\begin{align}
C^{\text{P2}}_{9,3} = -\frac{23 \pi^3 i}{2}.    
\end{align}
For constants at higher weights, the PSLQ algorithm can help to find the rational combinations.
We provide analytic results of all the canonical bases in an ancillary file along with this paper.

\section{Analytical NNLO QCD corrections}
\label{sec:nnlo}

Based on the analytical results of the master integrals,
we obtain the results for three-loop contributions to the cross section in eq. (\ref{eq:xsection}).
All the infrared divergences have been canceled. 
However, there are remaining ultraviolet divergences.
They cancel out after including the contributions from counter-terms.
The calculation of this part is standard and thus we refrain from giving the details here. The interested readers are referred to ref.~\cite{Wang:2023xud}.

As mentioned in section \ref{sec:frame}, there are two kinds of diagrams with different coupling structures contributing to the cross section,
i.e., 
\begin{align}
    X_{2} =  X_{2,Vt\bar{t}}+ 
     X_{2,gt\bar{t}}
\end{align}
with 
\begin{align}
     X_{2,Vt\bar{t}} & =  X_{2,Vt\bar{t}}^{ZZ}+  X_{2,Vt\bar{t}}^{\gamma Z}+ X_{2,Vt\bar{t}}^{\gamma\gamma}\nn \\
     &+ \theta(z-16) \left[X_{2,Vt\bar{t}}^{ZZ,4t}+  X_{2,Vt\bar{t}}^{\gamma Z,4t}+ X_{2,Vt\bar{t}}^{\gamma\gamma,4t} \right]\nn \\
     X_{2,gt\bar{t}} & = X_{2,gt\bar{t}}^{ZZ}+ X_{2,gt\bar{t}}^{\gamma Z}+X_{2,gt\bar{t}}^{\gamma\gamma}\,.
\end{align}
In the first equation, we split the contributions according to the number of top (anti-)quarks crossing the cuts.
The last term represents the contribution from four top (anti-)quarks, which is non-vanishing only for $z\geqslant 16\, (\sqrt{s}\geqslant 4m_t)$.
The full analytical results for all the above terms are provided in an ancillary file.
Below we present the results of maximal transcendental {\it weight} explicitly for $X_{2,Vt\bar{t}}$.
\begin{align}
&X_{2,Vt\Bar{t}}^{Z Z} = 
\frac{c_{\text{vec,u}}^2}{\left(c_{\text{axi}}^2 (w+1)^2+c_{\text{vec,u}}^2 \left(w^2-4 w+1\right)\right)}
\nonumber\\&\quad
\times\bigg(
\frac{4 \left(w^2+1\right)^2 \left(w^2-4 w+1\right)}{9 \left(w^2-1\right)^2}\Big[
100 i \pi  G(0,-1,-1,w)+76 i \pi  G(0,-1,1,w)
\nonumber\\&\quad
+76 i \pi  G(0,1,-1,w)+35 i \pi  G(0,1,1,w)-100 G(0,-1,-1,0,w)-28 G(0,-1,0,-1,w)
\nonumber\\&\quad
-14 G(0,-1,0,1,w)-76 G(0,-1,1,0,w)+128 G(0,0,-1,-1,w)+64 G(0,0,-1,1,w)
\nonumber\\&\quad
+64 G(0,0,1,-1,w)+32 G(0,0,1,1,w)-76 G(0,1,-1,0,w)+4 G(0,1,0,-1,w)
\nonumber\\&\quad
+2 G(0,1,0,1,w)-35 G(0,1,1,0,w)\Big]
\nonumber\\&\quad
+\frac{4 \left(8 w^4-85 w^3+166 w^2-85 w+8\right)}{27 \left(w^2-1\right)}\Big[
8{\pi}^2G(1,x_1,w)+12i{\pi}G(1,x_1,0,w)
\nonumber\\&\quad
+3i{\pi}G(1,x_1,1,w)+18G(1,x_1,0,-1,w)-12G(1,x_1,0,0,w)+9G(1,x_1,0,1,w)
\nonumber\\&\quad
-3G(1,x_1,1,0,w)+8{\pi}^2G(1,x_2,w)+12i{\pi}G(1,x_2,0,w)+3i{\pi}G(1,x_2,1,w)
\nonumber\\&\quad
+18G(1,x_2,0,-1,w)-12G(1,x_2,0,0,w)+9G(1,x_2,0,1,w)-3G(1,x_2,1,0,w)
\nonumber\\&\quad
-3i{\pi}G(1,0,1,w)-18G(1,0,0,-1,w)-9G(1,0,0,1,w)+3G(1,0,1,0,w)
\Big]
\nonumber\\&\quad
-\frac{2 \left(32 w^6-181 w^5+230 w^4-192 w^3-86 w^2-11 w+16\right)}{27 \left(w^2-1\right)^2}\Big[
8{\pi}^2G(0,x_1,w)
\nonumber\\&\quad
+12i{\pi}G(0,x_1,0,w)+3i{\pi}G(0,x_1,1,w)+18G(0,x_1,0,-1,w)-12G(0,x_1,0,0,w)
\nonumber\\&\quad
+9G(0,x_1,0,1,w)-3G(0,x_1,1,0,w)+8{\pi}^2G(0,x_2,w)+12i{\pi}G(0,x_2,0,w)
\nonumber\\&\quad
+3i{\pi}G(0,x_2,1,w)+18G(0,x_2,0,-1,w)-12G(0,x_2,0,0,w)+9G(0,x_2,0,1,w)
\nonumber\\&\quad
-3G(0,x_2,1,0,w)
\Big]-\frac{2 w \left(24 w^5-145 w^4+182 w^3-96 w^2-110 w+49\right)}{9 \left(w^2-1\right)^2}
\nonumber\\&\quad
\times\Big[2 G(0,0,0,-1,w)+G(0,0,0,1,w)\Big]
\nonumber\\&\quad
+\frac{4 \left(90 w^6-671 w^5+1174 w^4-872 w^3-520 w^2-201 w+128\right)}{9 \left(w^2-1\right)^2}
\nonumber\\&\quad
\times\Big[G(0,-1,0,0,w)-i \pi  G(0,-1,0,w)\Big]
\nonumber\\&\quad
+\frac{w \left(6 w^5-23 w^4+10 w^3-24 w^2+8 w-1\right)}{9 \left(w^2-1\right)^2}\Big[G(0,0,0,0,w)-i \pi  G(0,0,0,w)\Big]
\nonumber\\&\quad
+\frac{8 \left(24 w^6-45 w^5-120 w^4+32 w^3+96 w^2+77 w-32\right)}{9 \left(w^2-1\right)^2}
\nonumber\\&\quad
\times\Big[G(0,0,-1,0,w)-i \pi  G(0,0,-1,w)\Big]
\nonumber\\&\quad
+\frac{4 \left(47 w^6-111 w^5-168 w^4+36 w^3+141 w^2+147 w-56\right)}{9 \left(w^2-1\right)^2}
\nonumber\\&\quad
\times\Big[G(0,0,1,0,w)-i \pi  G(0,0,1,w)\Big]
\nonumber\\&\quad
+\frac{4 \left(w^4-4 w^3+4 w^2-4 w+1\right)}{27 \left(w^2-1\right)}\Big[5 \pi ^2 G(1,0,w)+6 i \pi  G(1,0,0,w)-6 G(1,0,0,0,w)\Big]
\nonumber\\&\quad
+\frac{2 \left(29 w^6-424 w^5+1075 w^4-840 w^3-445 w^2-416 w+181\right)}{9 \left(w^2-1\right)^2}
\nonumber\\&\quad
\times\Big[G(0,1,0,0,w)-i \pi  G(0,1,0,w)\Big]
\nonumber\\&\quad
-\frac{2 \pi ^2 \left(361 w^6-2689 w^5+4707 w^4-3488 w^3-2091 w^2-799 w+511\right) G(0,-1,w)}{27 \left(w^2-1\right)^2}
\nonumber\\&\quad
-\frac{2 \pi ^2 w \left(8 w^5-31 w^4+16 w^3-32 w^2+8 w-1\right) G(0,0,w)}{27 \left(w^2-1\right)^2}
\nonumber\\&\quad
-\frac{2 \pi ^2 \left(19 w^6-281 w^5+713 w^4-560 w^3-293 w^2-279 w+121\right) G(0,1,w)}{9 \left(w^2-1\right)^2}
\nonumber\\&\quad
-\frac{4 i}{9 \left(w^2-1\right)}\Big[\pi ^3 \left(w^4-4 w^3+4 w^2-4 w+1\right)
\nonumber\\&\quad
-3 i \left(8 w^4-85 w^3+166 w^2-85 w+8\right) \zeta (3)\Big]G(1,w)
\nonumber\\&\quad
-\frac{1}{27 \left(w^2-1\right)^2}\Big[6 \left(15 w^6-61 w^5+53 w^4-64 w^3-5 w^2-3 w+1\right) \zeta (3)
\nonumber\\&\quad
-i \pi ^3 w \left(10 w^5-39 w^4+22 w^3-40 w^2+8 w-1\right)\Big]G(0, w)
\nonumber\\&\quad
+\frac{\pi ^4 \left(231 w^6-873 w^5+285 w^4-848 w^3+351 w^2+25 w-19\right)}{1620 \left(w^2-1\right)^2}
\nonumber\\&\quad
+\frac{2 i \pi  \left(15 w^6-61 w^5+53 w^4-64 w^3-5 w^2-3 w+1\right) \zeta (3)}{9 \left(w^2-1\right)^2}
\bigg)
\nonumber\\&\quad
+\frac{c_{\text{axi}}^2}{\left(c_{\text{axi}}^2 (w+1)^2+c_{\text{vec,u}}^2 \left(w^2-4 w+1\right)\right)}
\nonumber\\&\quad
\times\bigg(\frac{4 \left(8 w^4-35 w^3+190 w^2-35 w+8\right)}{27 \left(w^2-1\right)}\Big[
8{\pi}^2G(1,x_1,w)+8{\pi}^2G(1,x_2,w)
\nonumber\\&\quad
-3i{\pi}G(1,0,1,w)+12i{\pi}G(1,x_1,0,w)+3i{\pi}G(1,x_1,1,w)+12i{\pi}G(1,x_2,0,w)
\nonumber\\&\quad
+3i{\pi}G(1,x_2,1,w)-18G(1,0,0,-1,w)-9G(1,0,0,1,w)+3G(1,0,1,0,w)
\nonumber\\&\quad
+18G(1,x_1,0,-1,w)-12G(1,x_1,0,0,w)+9G(1,x_1,0,1,w)-3G(1,x_1,1,0,w)
\nonumber\\&\quad
+18G(1,x_2,0,-1,w)-12G(1,x_2,0,0,w)+9G(1,x_2,0,1,w)-3G(1,x_2,1,0,w)\Big]
\nonumber\\&\quad
-\frac{2 \left(32 w^5-19 w^4+273 w^3-177 w^2+67 w+16\right)}{27 (w-1)^2 (w+1)}\Big[
8{\pi}^2G(0,x_1,w)+8{\pi}^2G(0,x_2,w)
\nonumber\\&\quad
+12i{\pi}G(0,x_1,0,w)+3i{\pi}G(0,x_1,1,w)+12i{\pi}G(0,x_2,0,w)+3i{\pi}G(0,x_2,1,w)
\nonumber\\&\quad
+18G(0,x_1,0,-1,w)-12G(0,x_1,0,0,w)+9G(0,x_1,0,1,w)-3G(0,x_1,1,0,w)
\nonumber\\&\quad
+18G(0,x_2,0,-1,w)-12G(0,x_2,0,0,w)+9G(0,x_2,0,1,w)-3G(0,x_2,1,0,w)
\Big]
\nonumber\\&\quad
+\frac{4 \left(w^2+1\right)^2}{9 (w-1)^2}\Big[
100i{\pi}G(0,-1,-1,w)+76i{\pi}G(0,-1,1,w)+76i{\pi}G(0,1,-1,w)
\nonumber\\&\quad
+35i{\pi}G(0,1,1,w)-100G(0,-1,-1,0,w)-28G(0,-1,0,-1,w)-14G(0,-1,0,1,w)
\nonumber\\&\quad
-76G(0,-1,1,0,w)+128G(0,0,-1,-1,w)+64G(0,0,-1,1,w)+64G(0,0,1,-1,w)
\nonumber\\&\quad
+32G(0,0,1,1,w)-76G(0,1,-1,0,w)+4G(0,1,0,-1,w)+2G(0,1,0,1,w)
\nonumber\\&\quad
-35G(0,1,1,0,w)\Big]-\frac{2 w \left(24 w^5-55 w^4+300 w^3-498 w^2+276 w-55\right)}{9 (w-1)^3 (w+1)}
\nonumber\\&\quad
\times\Big[2G(0,0,0,-1,w)+G(0,0,0,1,w)\Big]
\nonumber\\&\quad
+\frac{w \left(6 w^5-w^4+8 w^3-20 w^2+2 w-1\right)}{9 (w-1)^3 (w+1)}\Big[G(0,0,0,0,w)-i \pi  G(0,0,0,w))\Big]
\nonumber\\&\quad
+\frac{4 \left(w^4+2 w^3+4 w^2+2 w+1\right)}{27 \left(w^2-1\right)}\Big[5\pi^2G(1,0,w)+6i \pi G(1,0,0,w)-6G(1,0,0,0,w)\Big]
\nonumber\\&\quad
+\frac{8 \left(24 w^6+53 w^5-314 w^4+466 w^3-306 w^2+53 w+32\right)}{9 (w-1)^3 (w+1)}
\nonumber\\&\quad
\times\Big[G(0,0,-1,0,w)-i \pi  G(0,0,-1,w)\Big]
\nonumber\\&\quad
+\frac{2 \left(29 w^6-316 w^5+1873 w^4-2788 w^3+1663 w^2-316 w-181\right)}{9 (w-1)^3 (w+1)}
\nonumber\\&\quad
\times\Big[G(0,1,0,0,w)-i \pi  G(0,1,0,w)\Big]
\nonumber\\&\quad
+\frac{4 \left(47 w^6+79 w^5-476 w^4+693 w^3-467 w^2+79 w+56\right)}{9 (w-1)^3 (w+1)}
\nonumber\\&\quad
\times\Big[G(0,0,1,0,w)-i \pi  G(0,0,1,w)\Big]
\nonumber\\&\quad
+\frac{4 \left(90 w^6-313 w^5+1820 w^4-2776 w^3+1602 w^2-313 w-128\right)}{9 (w-1)^3 (w+1)}
\nonumber\\&\quad
\times\Big[G(0,-1,0,0,w)-i \pi  G(0,-1,0,w)\Big]
\nonumber\\&\quad
-\frac{2 \pi ^2 \left(361 w^6-1251 w^5+7279 w^4-11100 w^3+6407 w^2-1251 w-511\right)G(0,-1,w)}{27 (w-1)^3 (w+1)}
\nonumber\\&\quad
-\frac{2 \pi ^2 w \left(8 w^5-w^4+10 w^3-24 w^2+2 w-1\right) G(0,0,w)}{27 (w-1)^3 (w+1)}
\nonumber\\&\quad
-\frac{2 \pi ^2 \left(19 w^6-211 w^5+1249 w^4-1860 w^3+1109 w^2-211 w-121\right) G(0,1,w)}{9 (w-1)^3 (w+1)}
\nonumber\\&\quad
-\frac{4}{9 \left(w^2-1\right)}\Big[3 \left(8 w^4-35 w^3+190 w^2-35 w+8\right) \zeta (3)
\nonumber\\&\quad
+i \pi ^3 \left(w^4+2 w^3+4 w^2+2 w+1\right)\Big]G(1,w)
\nonumber\\&\quad
-\frac{1}{27 (w-1)^3 (w+1)}\Big[6 \left(15 w^6+w^5+13 w^4-20 w^3-3 w^2+w-1\right) \zeta (3)
\nonumber\\&\quad
-i \pi ^3 w \left(10 w^5-w^4+12 w^3-28 w^2+2 w-1\right)\Big]G(0,w)
\nonumber\\&\quad
+\frac{\pi ^4 \left(231 w^6-51 w^5+333 w^4-908 w^3+121 w^2-51 w+19\right)}{1620 (w-1)^3 (w+1)}
\nonumber\\&\quad
+\frac{2 i \pi  \left(15 w^6+w^5+13 w^4-20 w^3-3 w^2+w-1\right) \zeta (3)}{9 (w-1)^3 (w+1)}
\nonumber\\&\quad
+\frac{4 w^2}{135 \left(w^2-1\right)}\Big[900 \zeta (3) G(0,w)-30 i \pi ^3 G(0,w)+180 i \pi ^3 G(1,w)-360 \pi ^2 G(0,-1,w)
\nonumber\\&\quad
+30 \pi ^2 G(0,0,w)-1080 \pi ^2 G(0,1,w)-300 \pi ^2 G(1,0,w)-720 i \pi  G(0,-1,0,w)
\nonumber\\&\quad
+1440 i \pi  G(0,0,-1,w)+1080 i \pi  G(0,0,1,w)-1440 i \pi  G(0,1,0,w)-360 i \pi  G(1,0,0,w)
\nonumber\\&\quad
+720 G(0,-1,0,0,w)-1440 G(0,0,-1,0,w)-1080 G(0,0,1,0,w)+1440 G(0,1,0,0,w)
\nonumber\\&\quad
+360 G(1,0,0,0,w)-900 i \pi  \zeta (3)+7 \pi ^4\Big]
\bigg)+\cdots
\end{align}
with
\begin{align}
x_1 = \frac{1+\sqrt{3}i}{2},\quad x_2 = \frac{1-\sqrt{3}i}{2}\,.
\end{align}
Here the color factors $C_F = 4/3$, $C_A = 3$, $T_R = 1/2$ have been substituted.
The omitted terms have transcendental $weight$ lower than four. 
The results for the other combinations of electroweak boson propagators are related to the above result by
\begin{align}
X_{2,Vt\Bar{t}}^{\gamma\gamma}= X^{\gamma Z}_{2,Vt\Bar{t}} = X_{2,Vt\Bar{t}}^{Z Z}\bigg|_{c_{\text{axi}}=0}\,.
\end{align}

The contribution from the four top (anti-)quarks contains master integrals depending on elliptic integrals.
It was found that one can choose such a regular basis that only the $\mathcal{O}(\ep^0)$ parts of the master integrals are required \cite{Lee:2019wwn}.
The regular basis consists of twelve integrals $F_i^{4t}(z),i=1,\cdots,12,$ that can be expressed either as
complete elliptic integrals of the first kind or one-fold integrals of them.
Their explicit definitions are given in the appendix of \cite{Lee:2019wwn}.
Based on this basis, we give the result of 
four top quark production, 
\begin{align}
&X_{2,Vt\Bar{t}}^{ZZ,4t} = 
\frac{c_{\text{vec,u}}^2}{\left(c_{\text{axi}}^2 (z-4)+c_{\text{vec,u}}^2 \left(z+2\right)\right)}\bigg(
\frac{F^{4t}_1(z) \left(3 z^2+430 z+128\right)}{10368 \pi  r_1}
\nonumber\\&\quad
+\frac{F^{4t}_2(z) \left(85 z^3+990 z^2+16680 z-92608\right)}{15552 \pi  r_1 (z-4)}
-\frac{F^{4t}_3(z) \left(1019 z^2-1286 z-12120\right)}{3888 \pi  r_1}
\nonumber\\&\quad
-\frac{F^{4t}_4(z) (z-6) \left(49 z^2-336 z+464\right)}{1296 \pi  (z-4)^2}
-\frac{F^{4t}_5(z) (z-1) (z+2)}{108 \pi  r_1}
\nonumber\\&\quad
-\frac{F^{4t}_6(z)(z) \left(z^2+2 z-18\right)}{54 \pi  (z-4)}
+\frac{F^{4t}_7(z) \left(5 z^2-12 z-50\right)}{216 \pi  r_1}
\nonumber\\&\quad
+\frac{F^{4t}_8(z) \left(5 z^2-2 z-60\right)}{108 \pi  (z-4)}
+\frac{F^{4t}_9(z) \left(z^2+z+4\right)}{108 \pi  r_1}
\nonumber\\&\quad
+\frac{F^{4t}_{10}(z) \left(z^2-2\right)}{54 \pi  r_1}
-\frac{F^{4t}_{12}(z) (z+2) (z-2)^2}{54 \pi  (z-4) z}\bigg)
\nonumber\\&\quad+
\frac{c_{\text{axi}}^2}{\left(c_{\text{axi}}^2 (z-4)+c_{\text{vec,u}}^2 \left(z+2\right)\right)}
\bigg(-\frac{F_1^{4t}(z) \left(114 z^2+2189 z+2140\right)}{10368 \pi r_1}
\nonumber\\
&\quad+\frac{F^{4t}_2(z) \left(490 z^3+5821 z^2-21476 z+20256\right)}{15552 \pi  r_1 z}
-\frac{F^{4t}_3(z) r_1 \left(1424 z^2-2235 z+2532\right)}{3888 \pi  z^2}
\nonumber\\&\quad
-\frac{F^{4t}_4(z) \left(85 z^3-980 z^2+2076 z+1296\right)}{1296 \pi  (z-4) z}
-\frac{F^{4t}_5(z) \left(4 z^2-11 z+16\right)}{432 \pi  r_1}
\nonumber\\&\quad
-\frac{F^{4t}_6(z) \left(4 z^2-55 z+6\right)}{216 \pi  z}
+\frac{F^{4t}_7(z) \left(5 z^3-18 z^2+2 z+84\right)}{216 \pi  r_1 z}
\nonumber\\&\quad
+\frac{F^{4t}_8(z) \left(10 z^2-79 z+6\right)}{216 \pi  z}
+\frac{F^{4t}_9(z) \left(z^3-7 z^2-36 z-6\right)}{108 \pi  r_1 z}
\nonumber\\&\quad
+\frac{F^{4t}_{10}(z) \left(z^2-6 z-2\right)}{54 \pi  r_1}-\frac{F^{4t}_{12}(z) (z-2)^2}{54 \pi  z}
\bigg),
\end{align}
where the color factors $C_F = 4/3$, $C_A = 3$ have been substituted. 
The results for the other electroweak boson combinations are given by
\begin{align}
X_{2,Vt\Bar{t}}^{\gamma\gamma,4t}= X_{2,Vt\Bar{t}}^{\gamma Z,4t} = X_{2,Vt\Bar{t}}^{ZZ,4t}\bigg|_{c_{\text{axi}}=0}\,.
\end{align}
We have compared with  \cite{Lee:2019wwn} for the result of $X_{2,Vt\Bar{t}}^{\gamma\gamma,4t}$ and found full agreement \footnote{There is a typo in eq. (4.6) of \cite{Lee:2019wwn}, i.e., the denominator $5s$ should be changed to $2s$.}.

The full result of $X_{2,gt\Bar{t}}$ can be written as 
\begin{align}
X_{2,gt\Bar{t}}^{VV'} &= C_F F_{2,gt\Bar{t}}^{VV'} \tilde{X}_{2,gt\Bar{t}}
\end{align}
with
\begin{align}
F_{2,gt\Bar{t}}^{\gamma\gamma} &= -\frac{e_d^2 n_{l,d}+e_u^2 n_{l,u}}{(z+2) e_u^2},\nn\\
F_{2,gt\Bar{t}}^{\gamma Z} & =-\frac{c_{\text{vec,u}} e_u n_{l,u}+c_{\text{vec,d}} e_d n_{l,d}}{(z+2) c_{\text{vec,u}} e_u},\quad\nonumber\\
F_{2,gt\Bar{t}}^{ZZ} &= -\frac{c^2_{\text{axi}} n_l+c^2_{\text{vec,d}} n_{l,d}+c^2_{\text{vec,u}} n_{l,u}}{c_{\text{axi}}^2 (z-4)+c^2_{\text{vec,u}}(z+2)}.
\end{align}
Here $e_u$ ($e_d$) and $n_{l,u}$ ($n_{l,d}$) stand for the charge and number of light up-type (down-type) quarks, respectively, and 
$n_l=n_{l,u}+n_{l,d}$.
It can be seen that 
\begin{align}
F_{2,gt\Bar{t}}^{\gamma\gamma} = F_{2,gt\Bar{t}}^{\gamma Z} = F_{2,gt\Bar{t}}^{ZZ}\,\,\, \text{if}\,\, c_{\text{axi}} = 0\,\, {\rm and} \,\, \frac{c_{\text{vec,u}}}{c_{\text{vec,d}}} = \frac{e_u}{e_d}.
\end{align}
The complete form of $\tilde{X}_{2,gt\Bar{t}} $ reads
\begin{align}
&\tilde{X}_{2,gt\Bar{t}} = \frac{(r_1-z)}{r_1-z+4}\bigg(
-\frac{z^2-6}{18 z}\Big[2 \pi ^2 G(-1,y)-\pi ^2 G(0,y)-\pi ^2 G(1,y)-\pi ^2 G(-i,y)
\nonumber\\&\quad
-\pi ^2 G(i,y)+3 G(-1,-1,-1,y)+3 G(-1,-1,0,y)-3 G(-1,-1,1,y)
\nonumber\\&\quad
-9 G(-1,0,-1,y)-9 G(-1,0,0,y)+9 G(-1,0,1,y)-9 G(-1,1,-1,y)
\nonumber\\&\quad
-9 G(-1,1,0,y)+9 G(-1,1,1,y)+6 G(-1,-i,-1,y)+6 G(-1,-i,0,y)
\nonumber\\&\quad
-6 G(-1,-i,1,y)+6 G(-1,i,-1,y)+6 G(-1,i,0,y)-6 G(-1,i,1,y)
\nonumber\\&\quad
-9 G(0,-1,-1,y)-9 G(0,-1,0,y)+9 G(0,-1,1,y)+3 G(0,0,-1,y)
\nonumber\\&\quad
+3 G(0,0,0,y)-3 G(0,0,1,y)+3 G(0,1,-1,y)+3 G(0,1,0,y)-3 G(0,1,1,y)
\nonumber\\&\quad
+6 G(0,-i,-1,y)+6 G(0,-i,0,y)-6 G(0,-i,1,y)+6 G(0,i,-1,y)+6 G(0,i,0,y)
\nonumber\\&\quad
-6 G(0,i,1,y)-9 G(1,-1,-1,y)-9 G(1,-1,0,y)+9 G(1,-1,1,y)+3 G(1,0,-1,y)
\nonumber\\&\quad
+3 G(1,0,0,y)-3 G(1,0,1,y)+3 G(1,1,-1,y)+3 G(1,1,0,y)-3 G(1,1,1,y)
\nonumber\\&\quad
+6 G(1,-i,-1,y)+6 G(1,-i,0,y)-6 G(1,-i,1,y)+6 G(1,i,-1,y)+6 G(1,i,0,y)
\nonumber\\&\quad
-6 G(1,i,1,y)+6 G(-i,-1,-1,y)+6 G(-i,-1,0,y)-6 G(-i,-1,1,y)
\nonumber\\&\quad
+6 G(-i,0,-1,y)+6 G(-i,0,0,y)-6 G(-i,0,1,y)+6 G(-i,1,-1,y)+6 G(-i,1,0,y)
\nonumber\\&\quad
-6 G(-i,1,1,y)-12 G(-i,-i,-1,y)-12 G(-i,-i,0,y)+12 G(-i,-i,1,y)
\nonumber\\&\quad
-12 G(-i,i,-1,y)-12 G(-i,i,0,y)+12 G(-i,i,1,y)+6 G(i,-1,-1,y)
\nonumber\\&\quad
+6 G(i,-1,0,y)-6 G(i,-1,1,y)+6 G(i,0,-1,y)+6 G(i,0,0,y)-6 G(i,0,1,y)
\nonumber\\&\quad
+6 G(i,1,-1,y)+6 G(i,1,0,y)-6 G(i,1,1,y)-12 G(i,-i,-1,y)-12 G(i,-i,0,y)
\nonumber\\&\quad
+12 G(i,-i,1,y)-12 G(i,i,-1,y)-12 G(i,i,0,y)+12 G(i,i,1,y)\Big]
\nonumber\\&\quad
+\frac{(z+4) (19 z+46)}{18 r_2}\Big[G(-1,-1,y)+G(-1,0,y)-G(-1,1,y)-G(0,-1,y)
\nonumber\\&\quad
-G(0,0,y)+G(0,1,y)-G(1,-1,y)-G(1,0,y)+G(1,1,y)\Big]
\nonumber\\&\quad
-\frac{\left(19 z^2+72 z+72\right)}{108 z}\Big[3 i \pi  G(0,w)-6 i \pi  G(1,w)-3 G(0,0,w)+6 G(1,0,w)+2 \pi ^2\Big]
\nonumber\\&\quad
-\frac{\left(73 z^2+222 z+189\right)}{54 z}\Big[G(0,w)-i\pi\Big]
-\frac{2 (z-4) (2123 z+4978)}{1296 r_1}
\nonumber\\&\quad
+\frac{\pi ^2 \left(19 z^2+122 z+184\right)}{72 r_2}+\frac{5 \left(z^2-6\right) \zeta (3)}{6 z}
\bigg)\,.
\end{align}

\section{Asymptotic expansion}
\label{sec:exp}

With the full analytic results at hand, it is interesting to investigate the asymptotic expansions in some limits.

We first consider the threshold limit  when the quark velocity $\beta=\sqrt{1-4/z}$ is small.
The QCD NLO corrections near the threshold are 
\begin{align}
X_{1,Vt\Bar{t}}^{Z Z}|_{\beta\rightarrow 0 }&=
\frac{c_{\text{vec,u}}^2(z+2)C_F}{\left(c_{\text{axi}}^2 \left(z-4\right)+c_{\text{vec,u}}^2 \left(z+2\right)\right)}
\bigg(\frac{\pi^2}{2\beta}-4+\frac{\pi^2\beta}{2}+\mathcal{O}\left(\beta^2\right)\bigg)
\nonumber\\&\quad
+\frac{c_{\text{axi}}^2(z-4)C_F}{\left(c_{\text{axi}}^2 \left(z-4\right)+c_{\text{vec,u}}^2 \left(z+2\right)\right)}
\bigg(\frac{\pi^2}{2\beta}-2+\frac{\pi^2\beta}{2}+\mathcal{O}\left(\beta^2\right)\bigg)\,.
\end{align}
As in the full result, $X_{1,Vt\Bar{t}}^{\gamma \gamma}|_{\beta\rightarrow 0 }$ and $X_{1,Vt\Bar{t}}^{\gamma Z}|_{\beta\rightarrow 0 }$ can be obtained from $X_{1,Vt\Bar{t}}^{Z Z}|_{\beta\rightarrow 0 }$ by setting $c_{\text{axi}}=0$.
Notice that we do not expand the prefactors in $X_{1,Vt\Bar{t}}^{Z Z}|_{\beta\rightarrow 0 }$. They arise because the corrections are compared to the LO result which contains
\begin{align}
c_{\text{axi}}^2 \left(z-4\right)+c_{\text{vec,u}}^2 \left(z+2\right) = \frac{-4 c_\text{axi}^2 \beta ^2+2 \left(\beta ^2-3\right) c_\text{vec,u}^2}{\beta ^2-1} \,.
\label{eq:zzlo}
\end{align}
It is observed that the axial-vector current gives a contribution suppressed by $\beta^2$ relative to the vector current.

With the help of the {\tt PolyLogTools} package  \cite{Duhr:2019tlz} and the relations among MPLs in \cite{Henn:2015sem,Frellesvig:2016ske}, the QCD NNLO corrections near the threshold can be expanded as 
\begin{align}
&X_{2,Vt\Bar{t}}^{Z Z}|_{\beta\rightarrow0 } = 
\frac{c_{\text{vec,u}}^2(z+2)}{\left(c_{\text{axi}}^2 \left(z-4\right)+c_{\text{vec,u}}^2 \left(z+2\right)\right)}
\bigg(C_F^2\Big[\frac{\pi ^4}{12 \beta ^2}-\frac{2 \pi ^2}{\beta }-\frac{2\pi^2\log(\beta)}{3}
+\frac{\pi ^4}{6}
\nonumber\\&\quad
+\frac{4\pi ^2\log(2)}{3}
-\zeta (3)-\frac{35 \pi ^2}{18}
+\frac{39}{4}+\frac{8\pi ^2\log(\beta)}{3}\beta+\frac{4\pi ^2\log(2)}{3}\beta-\frac{55 \pi ^2}{18}\beta
+\mathcal{O}\left(\beta^2\right)
\Big]
\nonumber\\&\quad
+C_AC_F\Big[-\frac{11\pi ^2\log(\beta)}{12 \beta }+\frac{11\pi ^2}{24 \beta }\log\left(\frac{\mu^2}{s}\right)+\frac{31 \pi ^2}{72 \beta }
-\pi ^2 \log (\beta)-\frac{11}{3}\log\left(\frac{4\mu^2}{s}\right)
\nonumber\\&\quad
-\frac{8\pi ^2 \log(2)}{3}-\frac{13 \zeta (3)}{2}+\frac{179 \pi ^2}{72}-\frac{151}{36}
-\frac{\pi ^2}{4}\beta \log (\beta )+\frac{11\pi ^2 }{24}\beta \log\left(\frac{\mu^2}{s}\right)
\nonumber\\&\quad
+\frac{8\pi ^2 \log (2)\beta }{3}+\frac{193 \pi ^2\beta }{72}
+\mathcal{O}\left(\beta^2\right)
\Big]
\nonumber\\&\quad
+C_F n_l\Big[\frac{\pi ^2 \log (\beta )}{6 \beta }-\frac{\pi ^2}{12 \beta }\log\left(\frac{\mu^2}{s}\right)-\frac{5 \pi ^2}{36 \beta }
+\frac{2}{3}\log\left(\frac{4\mu^2}{s}\right)+\frac{11}{18}
+\frac{\pi ^2\beta\log(\beta)}{6}
\nonumber\\&\quad
-\frac{\pi ^2}{12}\beta \log\left(\frac{\mu^2}{s}\right)-\frac{13 \pi ^2\beta }{36}
+\mathcal{O}\left(\beta^2\right)
\Big]\nonumber\\&\quad
+C_F\Big[-\frac{\pi ^2}{12 \beta }\log\left(\frac{4\mu^2}{s}\right)+\frac{2}{3}\log\left(\frac{4\mu^2}{s}\right)-\frac{2 \pi ^2}{9}+\frac{22}{9}
-\frac{\pi ^2}{12}\beta\log\left(\frac{4\mu^2}{s}\right)
\nonumber\\&\quad
-\frac{\pi ^2\beta }{12}+\mathcal{O}\left(\beta^2\right)\Big]\bigg)
\nonumber\\&\quad
+\frac{c_{\text{axi}}^2(z-4)}{\left(c_{\text{axi}}^2 \left(z-4\right)+c_{\text{vec,u}}^2 \left(z+2\right)\right)}
\bigg(
C_F^2\Big[\frac{\pi ^4}{12 \beta ^2}+\frac{\pi ^2}{4 \beta ^2}-\frac{\pi ^2}{\beta }-\frac{5\pi ^2 }{6}\log(\beta )
+\frac{\pi ^4}{6}
\nonumber\\&\quad
+\frac{3\pi ^2 \log (2)}{4}
-\frac{27 \zeta (3)}{8}
+\frac{\pi ^2}{12}+\frac{35}{12}+\frac{8\pi ^2 }{3}\beta\log (\beta )+\frac{4\pi ^2\log (2)\beta}{3}
-\frac{32 \pi ^2 }{9}\beta+\mathcal{O}\left(\beta^2\right)\Big]
\nonumber\\&\quad
+C_AC_F\Big[-\frac{11 \pi ^2 \log (\beta )}{12 \beta }+\frac{11 \pi ^2}{24 \beta }\log\left(\frac{\mu^2}{s}\right)+\frac{97 \pi ^2}{72 \beta }
-\frac{\pi ^2\log (\beta )}{3}-\frac{11}{6}\log\left(\frac{4\mu^2}{s}\right)
\nonumber\\&\quad
-\frac{3\pi ^2 \log (2)}{2}-\frac{9 \zeta (3)}{4}+\frac{89 \pi ^2}{72}-\frac{101}{36}-\frac{\pi ^2 \beta  \log (\beta )}{4}
+\frac{11}{24} \pi ^2 \beta\log\left(\frac{\mu^2}{s}\right)
\nonumber\\&\quad
+\frac{37 \pi ^2 \beta }{24}+\frac{8\pi ^2\log (2)\beta}{3}+\mathcal{O}\left(\beta^2\right)\Big]
\nonumber\\&\quad
+C_F n_l\Big[\frac{\pi ^2 \log (\beta )}{6 \beta }-\frac{\pi ^2}{12 \beta }\log\left(\frac{\mu^2}{s}\right)
-\frac{11 \pi ^2}{36 \beta }+\frac{1}{3}\log\left(\frac{4\mu^2}{s}\right)+\frac{7}{18}
\nonumber\\&\quad
+\frac{\pi ^2 \beta  \log (\beta )}{6}-\frac{\pi ^2 \beta}{12}\log\left(\frac{\mu^2}{s}\right)-\frac{2 \pi ^2 \beta }{9}+\mathcal{O}\left(\beta^2\right)\Big]
\nonumber\\&\quad
+C_F\Big[-\frac{\pi ^2}{12 \beta }\log\left(\frac{4\mu^2}{s}\right)+\frac{1}{3}\log\left(\frac{\mu^2}{s}\right)-\frac{13 \pi ^2}{24}
+\frac{2 \log ^2(2)}{3}+\frac{2\pi ^2 \log (2)}{3}
\nonumber\\&\quad
-\frac{4 \log (2)}{3}+\frac{20}{9}-\frac{\pi ^2}{12}\beta\log\left(\frac{4\mu^2}{s}\right)-\frac{\pi ^2 \beta}{12}+\mathcal{O}\left(\beta^2\right)\Big]\bigg).
\end{align}
We have compared the expression of $X_{2,Vt\Bar{t}}^{ZZ}|_{\beta\rightarrow0 }$  with the two-loop virtual corrections near the threshold \cite{Czarnecki:1997vz,Bernreuther:2004th,Bernreuther:2005rw,Ablinger:2017hst}, finding full agreement for the terms up to $\mathcal{O}\left(\beta^0\right)$.

The fixed-order predictions for the cross sections near the threshold are not valid anymore because the contributions from the $1/\beta^2$ terms become divergent.
These terms are induced by the color Coulomb interaction between the final-state heavy quarks.
The resummation of the leading divergent terms of the form $\alpha_s^n/\beta^n$ gives rise to the well-known Sommerfeld-Sakharov factor.
The subleading divergent terms $\alpha_s^{n+1}/\beta^n$ can also be resummed to all orders by calculating the Coulomb Green function. See, e.g., refs. \cite{Beneke:1999qg} and \cite{Beneke:2013kia} for the S- and P-wave contributions, respectively.

We remark that the contribution from the diagrams in which the top quarks do not couple to the electroweak boson directly is strongly suppressed near the threshold,
\begin{align}
\tilde{X}_{2,gt\Bar{t}}|_{\beta\rightarrow0 } = -\frac{128}{45045}\beta^{12}+\mathcal{O}\left(\beta^{13}\right).
\end{align}
Therefore, it is reasonable to neglect this contribution near the threshold.

Then we consider the limit of $z\to \infty$,
which corresponds to taking the massless limit of the quark mass.
The QCD NLO corrections are expanded in this limit as
\begin{align}
X_{1,Vt\Bar{t}}^{ZZ}|_{z\rightarrow\infty } &= \frac{c_{\text{vec,u}}^2(z+2)C_F}{\left(c_{\text{axi}}^2 \left(z-4\right)+c_{\text{vec,u}}^2 \left(z+2\right)\right)}
\bigg(\frac{3}{4}+\frac{9}{z}+\mathcal{O}\left(\frac{1}{z^2}\right)\bigg)
\nonumber\\&\quad
+\frac{c_{\text{axi}}^2(z-4)C_F}{\left(c_{\text{axi}}^2 \left(z-4\right)+c_{\text{vec,u}}^2 \left(z+2\right)\right)}\bigg(\frac{3}{4}+\frac{9\log(z)}{z}+\mathcal{O}\left(\frac{1}{z^2}\right)\bigg),
\end{align}
and 
\begin{align}
X_{1,Vt\Bar{t}}^{\gamma\gamma}|_{z\rightarrow\infty } = X_{1,Vt\Bar{t}}^{\gamma Z}|_{z\rightarrow\infty } = C_F \bigg(\frac{3}{4}+\frac{9}{z}+\mathcal{O}\left(\frac{1}{z^2}\right)\bigg).
\end{align}
We see that the limit can be taken continuously, though there is a logarithmic enhancement at subleading power in the axial-vector coupling case.

At NNLO, the contributions from two and four top quark final states are singular individually, i.e.,
they are divergent as $z\to \infty$.
This is due to the infrared divergences in the massless limit.
A similar behavior has been observed in $H\to b\bar{b}$ \cite{Wang:2023xud}.
However, their sum is finite in this limit,
and is given by
\begin{align}
&\quad\quad (X_{2,Vt\Bar{t}}^{ZZ}+X_{2,Vt\Bar{t}}^{ZZ,4t})|_{z\rightarrow\infty } = \frac{c_{\text{vec,u}}^2(z+2)}{\left(c_{\text{axi}}^2 \left(z-4\right)+c_{\text{vec,u}}^2 \left(z+2\right)\right)}\times
\nonumber\\&\quad
\bigg(C_F^2\Big[
-\frac{3}{32}-\frac{27 \log(z)}{2 z} + \frac{9}{8 z}+\mathcal{O}\left(\frac{1}{z^2}\right)\Big]
\nonumber\\&\quad
+C_AC_F\Big[\frac{11}{16} \log \left(\frac{\mu ^2}{s}\right)+\frac{123}{32}-\frac{11 \zeta (3)}{4}+\frac{33}{4 z}\log \left(\frac{\mu ^2}{s}\right)
+\frac{185}{8 z}+\mathcal{O}\left(\frac{1}{z^2}\right)\Big]
\nonumber\\&\quad
+C_Fn_l\Big[
-\frac{1}{8} \log \left(\frac{\mu ^2}{s}\right)
+\frac{\zeta (3)}{2}-\frac{11}{16}-\frac{3}{2 z}\log \left(\frac{\mu ^2}{s}\right)-\frac{13}{4 z}+\mathcal{O}\left(\frac{1}{z^2}\right)
\Big]
\nonumber\\&\quad
+C_F\Big[
-\frac{1}{8} \log \left(\frac{\mu ^2}{s}\right)
+\frac{\zeta (3)}{2}-\frac{11}{16}-\frac{3}{2 z}\log \left(\frac{\mu ^2}{s}\right)-\frac{13}{4 z}+\mathcal{O}\left(\frac{1}{z^2}\right)
\Big]
\bigg)
\nonumber\\&\quad+
\frac{c_{\text{axi}}^2(z-4)}{\left(c_{\text{axi}}^2 \left(z-4\right)+c_{\text{vec,u}}^2 \left(z+2\right)\right)}\times
\nonumber\\&\quad
\bigg(
C_F^2\Big[-\frac{3}{32}-\frac{27 \log ^2(z)}{4z}+\frac{63 \log (z)}{8z}-\frac{6 \pi ^2 \log (2)}{z}
+\frac{45 \zeta (3)}{2z}+\frac{6 \pi ^2}{z}-\frac{195}{8z}+\mathcal{O}\left(\frac{1}{z^2}\right)\Big]
\nonumber\\&\quad
+C_AC_F\Big[\frac{11}{16}\log \left(\frac{\mu ^2}{s}\right)-\frac{11\zeta (3)}{4}+\frac{123}{32}+\frac{33 \log ^2(z)}{8z}+\frac{185 \log (z)}{8z}
+\frac{33}{4z} \log (z) \log \left(\frac{\mu ^2}{s}\right)
\nonumber\\&\quad
+\frac{3 \pi ^2 \log (2)}{z}
+\frac{9 \zeta (3)}{4z}+\frac{3 \pi ^2}{8z}-\frac{89}{16z}+\mathcal{O}\left(\frac{1}{z^2}\right)\Big]
\nonumber\\&\quad
+C_Fn_l\Big[-\frac{1}{8}\log \left(\frac{\mu ^2}{s}\right)+\frac{\zeta (3)}{2}-\frac{11}{16}-\frac{3 \log ^2(z)}{4z}
-\frac{3\log (z) }{2z}\log \left(\frac{\mu ^2}{s}\right)
\nonumber\\&\quad
-\frac{13 \log (z)}{4z}-\frac{3 \pi ^2}{4z}+\frac{7}{8z}
+\mathcal{O}\left(\frac{1}{z^2}\right)\Big]
\nonumber\\&\quad
+C_F\Big[-\frac{1}{8}\log \left(\frac{\mu ^2}{s}\right)+\frac{\zeta (3)}{2}
-\frac{\pi ^2}{12}
+\frac{1}{4}-\frac{5 \log ^2(z)}{4z}
-\frac{3\log (z)}{2z} \log \left(\frac{\mu ^2}{s}\right)
\nonumber\\&\quad
-\frac{25 \log (z)}{4z}+\frac{13\pi^2}{12z}+\frac{1}{z}
+\mathcal{O}\left(\frac{1}{z^2}\right)\Big]\bigg)\,.
\end{align}
Again, the results for the $\gamma\gamma$ and $\gamma Z$ combinations can be obtained by setting $c_{\text{axi}} = 0$.
The contribution in this limit agrees with the results given in refs. \cite{Chetyrkin:1996cf,Chetyrkin:1997qi,Harlander:1997kw}.

Finally, the contribution of the processes in which the top quark pair is produced from a gluon is singular and has the following expansion form.
\begin{align}
&\quad\quad\tilde{X}_{2,gt\Bar{t}}|_{z\rightarrow \infty } 
\nonumber\\&\quad
= -\frac{z\log ^3(z)}{36}
+\frac{19z\log ^2(z)}{72}+\frac{\pi ^2z \log (z)}{18}-\frac{73z \log (z)}{54}
-\frac{5 \zeta (3)z}{6}-\frac{19 \pi ^2z}{108}+\frac{2123z}{648}
\nonumber\\&\quad
-\frac{\log ^3(z)}{18}
+\frac{55\log ^2(z)}{36}+\frac{\pi ^2\log (z)}{9}-\frac{127\log (z)}{27}
-\frac{5 \zeta (3)}{3}-\frac{55\pi ^2}{54}+\frac{4715}{324}+\mathcal{O}\left(\frac{1}{z^2}\right).
\label{eq:X2gtt}
\end{align}
This singular structure indicates that the cross section for a quark pair production with a specific flavor is sensitive to infrared effects, e.g., a gluon splitting to such a collinear quark pair.
Since the specific quark cannot be distinguished from the other massless quarks in this limit, it would be difficult to measure such a process with physical detectors.
A more realistic observable is the jet cross section which includes contributions from both the specific quark and other partons.
In this case, it is necessary to consider the diagrams with a top quark loop in the virtual gluon propagator.
Their contribution would cancel against the large logarithmic terms in eq.(\ref{eq:X2gtt}), giving rise to a finite result in the $z\to \infty$ limit.

\section{Numerical results}
\label{sec:num}

With the analytical results at hand, it is ready to present their numerical impact in phenomenological studies.
We take the numerical inputs as in ref. \cite{Chen:2022vzo},
\begin{align}
m_W &= 80.385 ~{\rm GeV},\quad m_Z = 91.1876  ~{\rm GeV},\quad m_t = 172.69 ~{\rm GeV},\quad\nonumber\\
\alpha &= \frac{1}{132.2},\quad \alpha_s(m_Z) = 0.1181,\quad \Gamma_Z = 2.4952~{\rm GeV},
\end{align}
and $\cos\theta_W = m_W/m_Z$. The package {\tt RunDec} \cite{Chetyrkin:2000yt,Herren:2017osy} was used to calculate $\alpha_s$ at other scales. 
First, we present the numerical values for the perturbative expansion coefficients at different orders in table \ref{X1X2}. From the LO results, one can see that $\sigma^{\gamma \gamma}$ is the dominant contribution to the $e^- e^+\to t\bar{t}$ process. 
The NLO coefficients, $X_1^{\gamma\gamma}$ and $X_1^{ZZ}$, decrease with the increasing of the collision energy.
Because of the axial current contribution, the corrections induced by a $Z$ boson exchange are larger than those by a photon.
The NNLO coefficients $X_{2,Vt\bar{t}}^{\gamma\gamma}$ and $X_{2,Vt\bar{t}}^{ZZ}$ exhibit similar features.
Moreover, it is interesting to note the large magnitude of these coefficients at $\sqrt{s}=350$ GeV.
Even multiplied by the expansion parameter $(\alpha_s/\pi)^i,i=1,2$, they provide corrections of about $100\%$ to the LO result.
Adopting the result with the Coulomb resummation is more appropriate in this threshold region as in \cite{Beneke:2015kwa,Beneke:2016kkb}.
All of the numerical results in table \ref{X1X2} have been cross checked with those in \cite{Chen:2022vzo} except for $X_{2,Vt\bar{t}}^{4t}$ 
and perfect agreements were found. 
We also compare the results of 
 $X_{2,Vt\bar{t}}^{\gamma\gamma}$ with the Pad\'e approximations in \cite{Maier:2017ypu} at
$\sqrt{s}$= 350, 500, 1000 GeV and $\mu = m_t$, finding the relative differences to be $\mathcal{O}(10^{-10})$, $\mathcal{O}(10^{-13})$ and $\mathcal{O}(10^{-28})$, respectively.

In $X_{2,Vt\bar{t}}^{ZZ}$, the contribution from flavour singlet diagrams is a special part that involves a quark loop.
It turns out to be a small correction,
only at most $2\%$ of the total $X_{2,Vt\bar{t}}^{ZZ}$ at $\sqrt{s}=1000 $ GeV.
The contributions of the four top quark final states 
or of the diagrams with a gluon splitting to two top quarks are also so tiny that they could be neglected in phenomenological studies.

\begin{table}[tb]
	\centering
	\scalebox{1.0}{
		\begin{tabular}{llll}
			\toprule
			$\quad\sqrt{s}$ &
			$\quad 350 $ GeV \quad &
			$\quad 500 $ GeV \quad &
			$\quad1000 $ GeV \quad \\
   			\midrule
			$\sigma^{\gamma \gamma}_{\text{LO}}~(\text{pb})$ &
			$ 2.4459 \times10^{-1}$ &
			$ 4.4577 \times10^{-1}$ &
			$ 1.2374 \times10^{-1}$ \\
			\midrule
			$\sigma^{Z Z}_{\text{LO}}~(\text{pb})$ &
			$1.5198\times10^{-2}$ &
			$8.2932\times10^{-2}$ &
			$3.7130\times10^{-2}$ \\
         	\midrule
          	$\sigma^{\gamma Z}_{\text{LO}}~(\text{pb})$ &
			$1.2490\times10^{-2}$ &
			$2.1948\times10^{-2}$ &
			$5.9393\times10^{-3}$ \\
         	\midrule
			$X_1^{\gamma\gamma}$&
			$35.87815637$ &
			$3.902679560$ &
			$1.454200325$ \\
			\midrule
			$X_1^{ZZ}$&
			$36.13473849$&
			$5.385227064$ &
			$2.242695157$ \\
			\midrule
			$X_{2,Vt\bar{t}}^{\gamma\gamma}$&
			$739.8408382$ &
			$15.48165668$ &
			$1.237511900$ \\
			\midrule
			$X_{2,Vt\bar{t}}^{ZZ}$&
			$779.1581353$ &
			$46.69192781$ &
			$11.22389311$ \\
   		\midrule
            $X_{2,Vt\bar{t}}^{ZZ,S}$&
            $3.524878161\times10^{-2}$&
            $-2.47153096\times10^{-1}$&
            $-2.09396980\times10^{-1}$\\   
            \midrule
            $X_{2,Vt\bar{t}}^{\gamma\gamma,4t}$&
			$0$ &
			$0$ &
			$4.852649148\times10^{-3}$ \\
                \midrule
                $X_{2,Vt\bar{t}}^{ZZ,4t}$&
			$0$ &
			$0$ &
			$2.389195068\times10^{-3}$ \\
   			\midrule
                $X_{2,gt\bar{t}}^{\gamma\gamma}$&
			$6.063662743\times10^{-13}$ &
			$2.166216155\times10^{-4}$ &
			$6.792148588\times10^{-2}$ \\
                \midrule
                $X_{2,gt\bar{t}}^{\gamma Z}$&
			$1.014082252\times10^{-13}$ &
			$3.622763087\times10^{-4}$ &
			$1.135913659\times10^{-1}$ \\
                \midrule
                $X_{2,gt\bar{t}}^{ZZ}$&
			$8.244429044\times10^{-12}$ &
			$9.145040414\times10^{-4}$ &
			$1.689525260\times10^{-1}$ \\
			\bottomrule
		\end{tabular}
  }
	\caption{Numerical results for the perturbative expansion coefficients at different collision energies. 
 $X_{2,Vt\bar{t}}^{ZZ,S}$ represents the contribution from the flavor singlet diagrams in $X_{2,Vt\bar{t}}^{ZZ}$. The renormalization scale is chosen as $\mu = \sqrt{s}$. } 
	\label{X1X2}
\end{table} 

\begin{figure}[ht]
		\centering
  \includegraphics[width=0.85\linewidth]{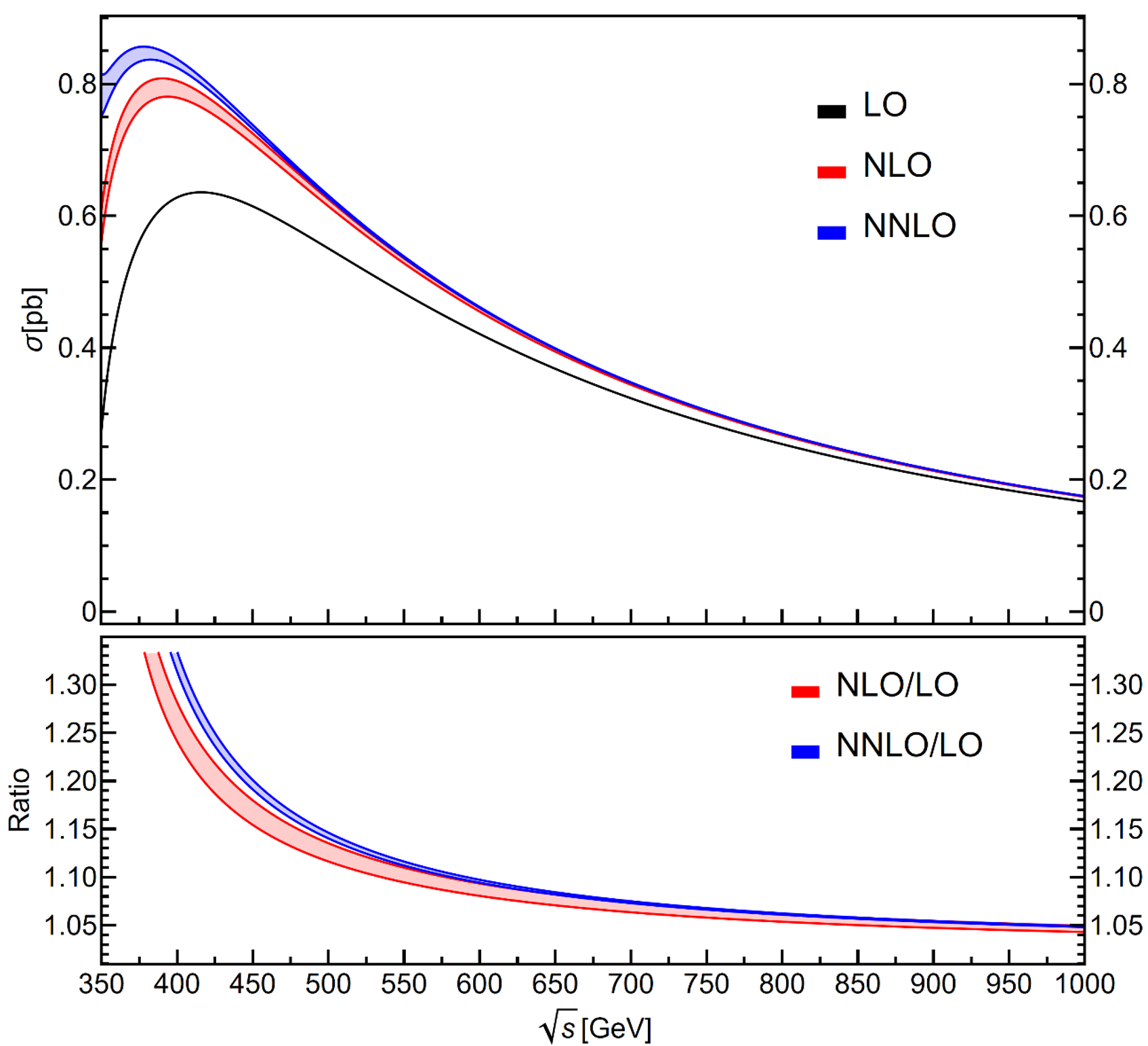}
\caption{Numerical results for the total cross sections of top quark pair production. The bands denote the scale uncertainties with the scale $\mu\in\left[\sqrt{s}/2,\,2\sqrt{s}\right]$. In the bottom plot, we show the ratios of higher-order results over the LO one. }
\label{fig:totalxs}
\end{figure}

Then we show the total cross sections of top quark pair production at different perturbative orders in figure \ref{fig:totalxs}.
The cross section first increases and then decreases when the collider energy changes from 350 GeV to 1000 GeV.
The peak positions move from 416 GeV at LO, to 
392 GeV at NLO and 381 GeV at NNLO at $\mu = \sqrt{s}$.
The higher-order QCD corrections are more significant for the process with smaller collision energies.
At $\sqrt{s}=400$ GeV, the NLO correction improves the LO cross section by $26\%$,
and the NNLO correction provides additional $5\%$ enhancement when choosing the renormalization scale $\mu = \sqrt{s}$.
At $\sqrt{s}=1000$ GeV, the NLO and  NNLO corrections reduce to only $5\%$  and $0.3\%$, respectively.
It is also important to observe that the QCD renormalization scale uncertainty, which is evaluated by varying the default scale $\mu=\sqrt{s}$ by a factor of two, 
is decreased from $3.2\%$ ($0.6\%$)  at NLO to $1.6\%$ ($0.09\%$) at NNLO for collisions at $\sqrt{s}=400$ (1000) GeV. 
Note that our results may not be valid near the top quark pair production threshold due to the reason discussed in the last section.

\begin{figure}[ht]
		\centering
		\includegraphics[width=0.8\linewidth]{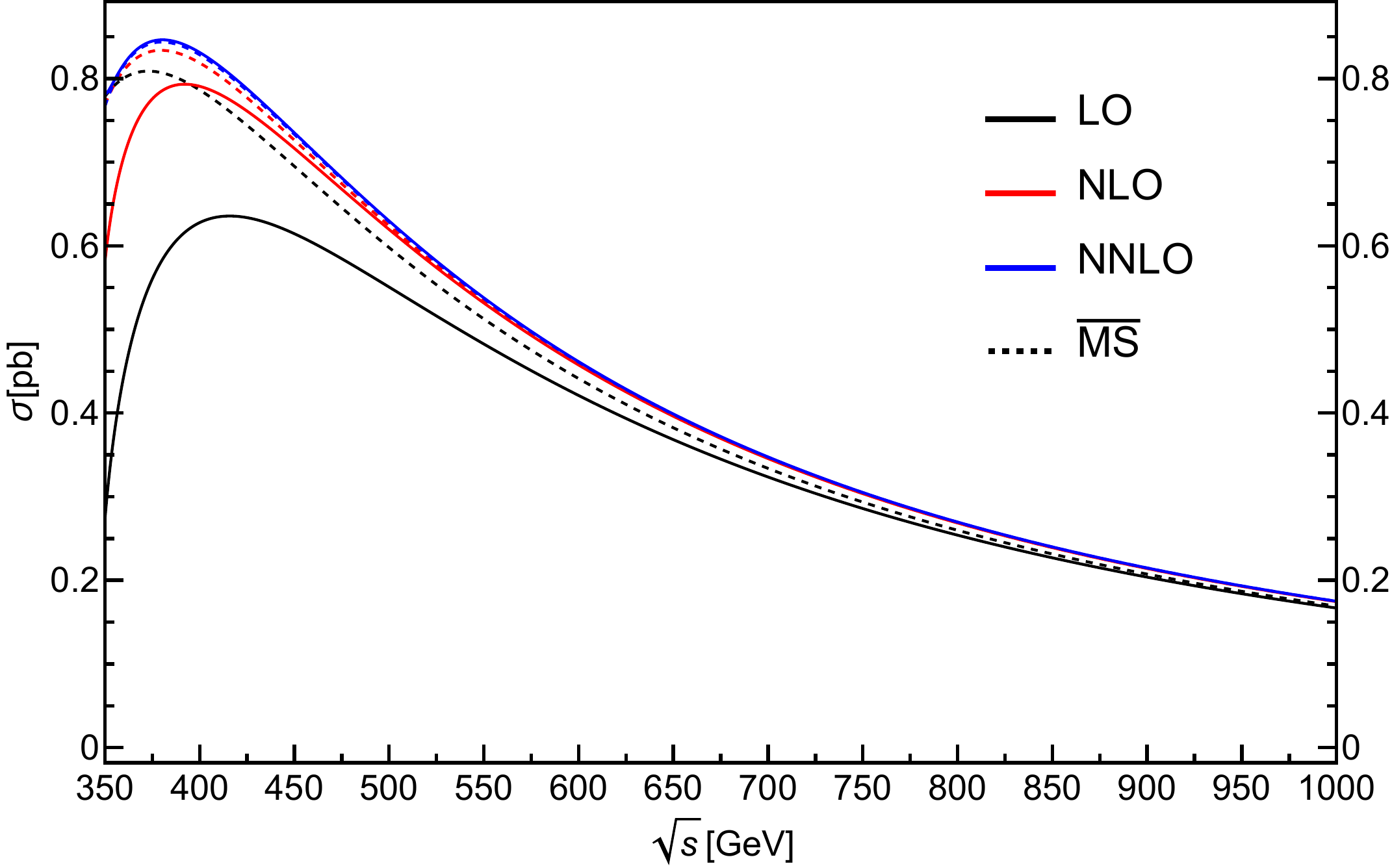}
\caption{The cross sections in the on-shell (solid lines) and $\overline{\rm MS}$ (dashed lines) renormalization schemes for the top quark mass. The renormalization scale $\mu = \sqrt{s}$ is used. }
\label{fig:scheme}
\end{figure}

We also provide the results in the $\overline{\rm MS}$ renormalization scheme which are converted from the on-shell scheme with the mass relation in \cite{Broadhurst:1991fy,Gray:1990yh,Marquard:2015qpa,Marquard:2016dcn}.
As shown in figure \ref{fig:scheme}, 
the cross sections in the $\overline{\rm MS}$ scheme exhibit much better convergence behavior.
The reason is that the pole mass is by definition plagued by the renormalon problem \cite{Beneke:1998ui}.
When more perturbative corrections are included, 
the difference between the two renormalization schemes is notably reduced.

\begin{figure}[ht]
		\centering
		\includegraphics[width=0.82\linewidth]{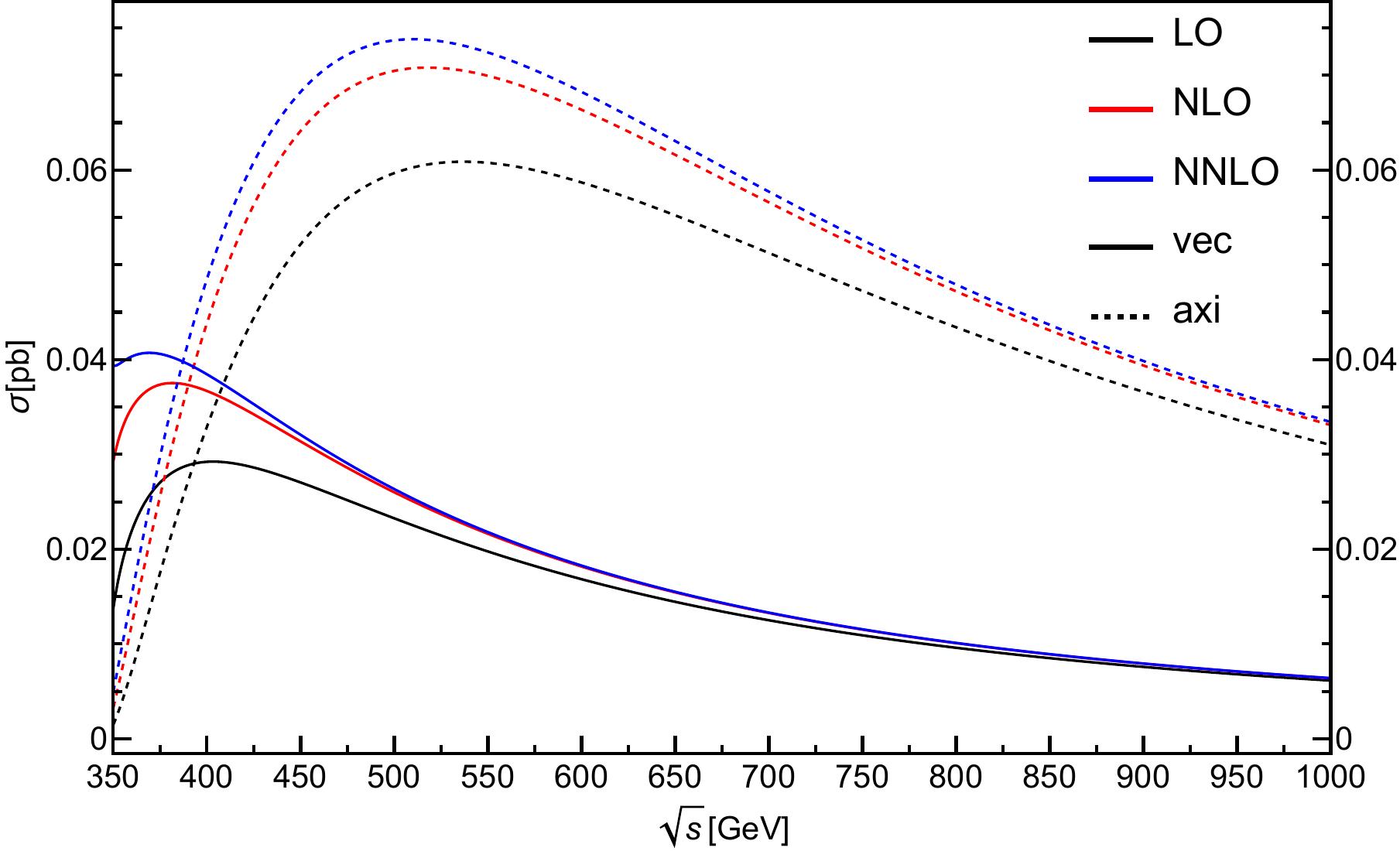}
\caption{The cross section $\sigma_{e^- e^+\to t\bar{t}}^{ZZ}$ as a function of $\sqrt{s}$.
The renormalization scale is chosen at $\mu = \sqrt{s}$. The solid and dash lines represent the vector current and axial-vector current contributions in $\sigma_{e^- e^+\to t\bar{t}}^{ZZ}$, respectively.}
\label{fig:axi-vec}
\end{figure}

One of the motivations for studying top quark pair production at electron colliders is to investigate the coupling between the top quarks and the $Z$ boson.
We show the cross section induced by $Z$ propagators $\sigma_{e^- e^+\to t\bar{t}}^{ZZ}$ in figure \ref{fig:axi-vec}. 
We split the contributions into the vector and axial-vector quark currents.
The axial-vector current contributions are strongly suppressed near the threshold; see eq.(\ref{eq:zzlo}) above.
With the increasing of the collision energy, the axial-vector contributions grow rapidly 
and become larger than the vector contributions when $\sqrt{s}>400$ GeV.
This is because the axial-vector current coupling $c^2_{\text{axi}}\,(\approx0.08)$ is larger than the vector current coupling $c^2_{\text{vec,u}}\,(\approx0.01)$.
We also see that the QCD corrections in the axial-vector 
 current contribution are more significant than those in the vector current.

\section{Conclusion}
\label{sec:conclusion}

In this paper, we have presented the analytic results of NNLO QCD corrections to top quark pair production at lepton colliders.
We consider the contributions from both a photon and a $Z$ boson propagator.
Part of the required master integrals are available in our previous paper on the $H\to bb$ decay,
and the others are computed explicitly using the canonical differential equations.
The full analytical NNLO QCD corrections are divided into the contributions of two and four top quark final states.
The former is expressed in terms of multiple polylogarithms, while the latter is written as a linear combination of complete elliptic integrals and one-fold integrals of them.
Near the threshold, the expansion up to $\mathcal{O}(\beta^0)$ agrees with the corresponding virtual corrections
and are compatible with the prediction of the Coulomb resummation formula. 
In the high energy limit, the sum of the contributions from the two and four top quark final states is finite, though they are individually divergent.
Numerically, the photon-induced process has a larger contribution to the total cross section than the $Z$ boson-induced one.
But the higher-order correction is more significant in the $Z$ boson-induced process.
At $\sqrt{s}=400$ GeV, the NNLO corrections increase the NLO result by $5\%$ 
and decrease the scale uncertainty by a factor of two.
The QCD corrections become smaller at higher collision energies.
We also find that the results in the $\overline{\rm MS}$ top quark mass scheme show faster perturbative convergence.
Numerical results explicitly illustrate that the axial-vector current contribution is suppressed near the threshold but enhanced at high energies compared to the vector current contribution.
Our analytical results are encoded in a Mathematica file that can be conveniently used for precise evaluation of the top quark pair production cross section.

\section*{Acknowledgements}

We would like to thank Long Chen, Xiang Chen, Xin Guan and Hai Tao Li for helpful discussions. This work was supported in part by the National Science Foundation of China (grant Nos. 12005117, 12175048, 12321005, 12375076) and the Taishan Scholar Foundation of Shandong province (tsqn201909011).
The topology diagrams in this paper were drawn using the TikZ-Feynman package \cite{Ellis:2016jkw}.

\appendix

\section{Topologies of the master integrals}
\label{appendix:topoMIs}

The topology diagrams of the master integrals in the P2 and P3 families
are displayed below.
\begin{figure}[ht]
	\centering
	\begin{minipage}{0.2\linewidth}
		\centering
		\includegraphics[width=1\linewidth]{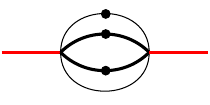}
		\caption*{$\M^{\text{P2}}_{1}$}
	\end{minipage}
	\begin{minipage}{0.2\linewidth}
		\centering
		\includegraphics[width=1\linewidth]{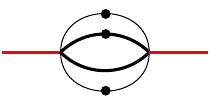}
		\caption*{$\M^{\text{P2}}_{2}$}
	\end{minipage}
	\begin{minipage}{0.2\linewidth}
		\centering
		\includegraphics[width=1\linewidth]{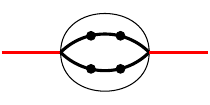}
		\caption*{$\M^{\text{P2}}_{3}$}
	\end{minipage}
	\centering
	\begin{minipage}{0.2\linewidth}
		\centering
		\includegraphics[width=1\linewidth]{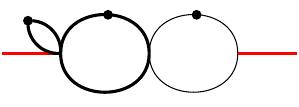}
		\caption*{$\M^{\text{P2}}_{4}$}
	\end{minipage}
	\begin{minipage}{0.2\linewidth}
		\centering
		\includegraphics[width=1\linewidth]{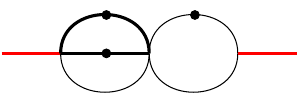}
		\caption*{$\M^{\text{P2}}_{5}$}
	\end{minipage}
	\begin{minipage}{0.2\linewidth}
		\centering
		\includegraphics[width=1\linewidth]{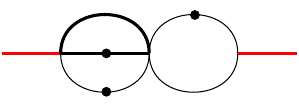}
		\caption*{$\M^{\text{P2}}_{6}$}
	\end{minipage}
	\begin{minipage}{0.17\linewidth}
		\centering
		\includegraphics[width=1\linewidth]{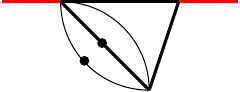}
		\caption*{$\M^{\text{P2}}_{7}$}
	\end{minipage}
	\begin{minipage}{0.2\linewidth}
		\centering
		\includegraphics[width=1\linewidth]{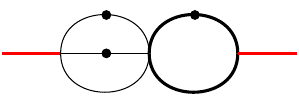}
		\caption*{$\M^{\text{P2}}_{8}$}
	\end{minipage}
	\begin{minipage}{0.2\linewidth}
		\centering
		\includegraphics[width=1\linewidth]{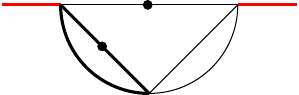}
		\caption*{$\M^{\text{P2}}_{9}$}
	\end{minipage}
	\begin{minipage}{0.2\linewidth}
		\centering
		\includegraphics[width=1\linewidth]{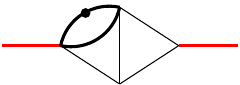}
		\caption*{$\M^{\text{P2}}_{10}$}
	\end{minipage}
	\begin{minipage}{0.2\linewidth}
		\centering
		\includegraphics[width=1\linewidth]{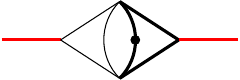}
		\caption*{$\M^{\text{P2}}_{11}$}
	\end{minipage}
	\begin{minipage}{0.2\linewidth}
		\centering
		\includegraphics[width=1\linewidth]{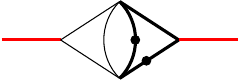}
		\caption*{$\M^{\text{P2}}_{12}$}
	\end{minipage}
	\begin{minipage}{0.2\linewidth}
		\centering
		\includegraphics[width=1\linewidth]{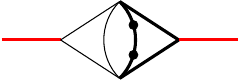}
		\caption*{$\M^{\text{P2}}_{13}$}
	\end{minipage}
	\begin{minipage}{0.2\linewidth}
		\centering
		\includegraphics[width=1\linewidth]{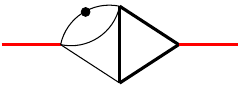}
		\caption*{$\M^{\text{P2}}_{14}$}
	\end{minipage}
	\begin{minipage}{0.28\linewidth}
		\centering
		\includegraphics[width=1\linewidth]{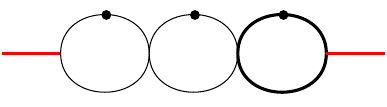}
		\caption*{$\M^{\text{P2}}_{15}$}
	\end{minipage}
	\begin{minipage}{0.24\linewidth}
		\centering
		\includegraphics[width=1\linewidth]{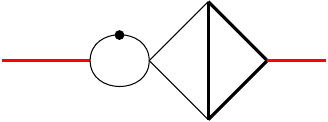}
		\caption*{$\M^{\text{P2}}_{16}$}
	\end{minipage}
\caption{Master integrals in the P2 family. The thick black and red lines stand for the massive top quark and the vector boson, respectively. One black dot indicates one additional power of the corresponding propagator.}
\label{P2_Topo}
\end{figure}
\begin{figure}[H]
	\centering
	\begin{minipage}{0.25\linewidth}
		\centering
		\includegraphics[width=0.8\linewidth]{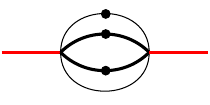}
		\caption*{$\M^{\text{P3}}_{1}$}
	\end{minipage}
	\begin{minipage}{0.25\linewidth}
		\centering
		\includegraphics[width=0.8\linewidth]{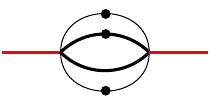}
		\caption*{$\M^{\text{P3}}_{2}$}
	\end{minipage}
	\begin{minipage}{0.25\linewidth}
		\centering
		\includegraphics[width=0.8\linewidth]{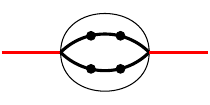}
		\caption*{$\M^{\text{P3}}_{3}$}
	\end{minipage}
	\centering
	\begin{minipage}{0.25\linewidth}
		\centering
		\includegraphics[width=1.1\linewidth]{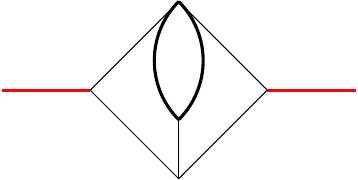}
		\caption*{$\M^{\text{P3}}_{4}$}
	\end{minipage}
 \quad\quad\quad
	\begin{minipage}{0.25\linewidth}
		\centering
		\includegraphics[width=1.1\linewidth]{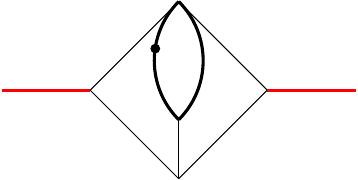}
		\caption*{$\M^{\text{P3}}_{5}$}
	\end{minipage}
\caption{Master integrals in the P3 family. The thick black and red lines stand for the massive top quark and the vector boson, respectively. One black dot indicates one additional power of the corresponding propagator.}
\label{P3_Topo}
\end{figure}
\bibliographystyle{JHEP}
\bibliography{reference}

\end{document}